\documentclass[preprint,amsmath,amssymb,aps,pre]{revtex4-2}

\usepackage{graphicx}
\usepackage{dcolumn}
\usepackage{bm}

\usepackage{accents}
\usepackage{color}
\allowdisplaybreaks

\usepackage{placeins}
\FloatBarrier

\begin{document}

\preprint{APS/123-QED}

\title{Dynamical mean field approach to associative memory model with non-monotonic transfer functions}

\author{Yoshiyuki Kabashima}
\email{kaba@phys.s.u-tokyo.ac.jp}
 \affiliation{Institute for Physics of Intelligence, The University of Tokyo, 7-3-1 Hongo, Bunkyo-ku, Tokyo 113-0033, Japan. \\
 Trans-Scale Quantum Science Institute, The University of Tokyo, 7-3-1 Hongo, Tokyo 113-0033, Japan.
 }


\author{Kazushi Mimura}
\email{mimura@hiroshima-cu.ac.jp}
\affiliation{%
 Department of Intelligent Systems, Hiroshima City University, 3-4-1 Ozuka-higashi, Asaminami-ku, Hiroshima 731-3194, Japan.
}%



\date{\today}

\begin{abstract}
The Hopfield associative memory model stores random patterns in synaptic couplings according to Hebb's rule and retrieves them through gradient descent on an energy function. This conventional setting, where neurons are assumed to have monotonic transfer functions, has been central to understanding associative memory. Morita (1993, Neural Netw. 6 115), however, showed that introducing non-monotonic transfer functions can dramatically enhance retrieval performance. While this phenomenon has been qualitatively examined, a full quantitative theory remains elusive due to the difficulty of analysis in the absence of an underlying energy function.
In this work, we apply dynamical mean-field theory to the discrete-time synchronous retrieval dynamics of the non-monotonic model, which succeeds in accurately characterizing its macroscopic dynamical properties. 
We also derive conditions for retrieval states, and clarify their relation to previous studies. Our results provide new insights into the non-equilibrium retrieval dynamics of associative memory models.
\end{abstract}

\maketitle


\section{\label{sec:introduction}Introduction}
The concept of associative memory in neural networks traces back to pioneering works in the early 1970s by Nakano~\cite{Nakano1972}, Amari~\cite{Amari1972}, Kohonen~\cite{Kohonen1972}, and others, who proposed models capable of storing and retrieving patterns through distributed representations. While these early studies laid the foundations of associative memory, the subsequent contribution of Hopfield (1982) was to reorganize and reformulate these ideas into a unified framework based on an energy function~\cite{Hopfield1982}. In the Hopfield model, random patterns are embedded in synaptic couplings according to Hebb’s rule, and retrieval proceeds via gradient descent dynamics on the associated energy landscape. This reformulation not only clarified the connection between neural networks and statistical physics but also provided a tractable paradigm for studying equilibrium properties using methods of equilibrium statistical mechanics~\cite{Amit1985,Nakanishi1997}.

Within this conventional framework, the neuronal transfer function is usually assumed to increase monotonically with input strength, under which the retrieval properties of the model are well understood. Morita (1993), however, demonstrated that introducing non-monotonic transfer functions can dramatically enhance retrieval performance, revealing an intriguing direction beyond the standard energy-based framework~\cite{Morita1993}. Building on this result, Yoshizawa, Morita, and Amari (1993) analyzed a piecewise-linear approximation of Morita’s model and showed that the storage capacity increases to nearly three times that of the Hopfield model, while spurious memories largely disappear~\cite{yoshizawa1993}. They also found, however, that the basin of attraction is not faithfully reproduced, highlighting the inherent difficulty in quantitatively characterizing retrieval dynamics in such systems.

In this work, our objective is to address this difficulty by applying the dynamical mean-field theory (DMFT), also known as the generating function method or the path-integral method~\cite{Sompolinsky1981,Eissfeller1992,Mimura2004,Mimura2014}. A key advantage of DMFT is that, by representing the trajectories of the dynamics directly in terms of generating functions, it enables direct analysis of the system’s dynamical behavior regardless of whether an energy function exists.
The Morita model was originally proposed for continuous-time retrieval dynamics. 
However, because DMFT requires a self-consistent determination of both the noise correlation function and the response function of the system components, its application to investigating long term behavior of continuous-time dynamics becomes computationally prohibitive~\cite{Faugeras2009,Kadmon2015,Zou2024}. For this reason, in the present study we focus on the discrete-time synchronous dynamics corresponding to Morita’s original continuous-time model. 
According to our analysis, when the time step size is set to $0.1$, the storage capacity, defined as the maximal number of retrievable random patterns $p_{\rm c}$ for a system of $N$ neurons, reaches approximately $p_{\rm c} \simeq 0.36N$. In addition, the evaluation of the basin of attraction agrees very well with numerical experiments for systems as large as $N=4096$.
Our analysis also revealed that the success or failure of retrieval manifests in qualitatively distinct behaviors of the system’s feedback coefficients, which correspond to Onsager’s reaction term in equilibrium theory~\cite{Thouless1977,Opper2016}. 
Further, we derived the conditions that must be satisfied by stationary states and discussed their relationship to equilibrium analyses presented in previous studies~\cite{Shiino1993,Waugh1993}.

The structure of this paper is as follows.
In Section 2, we define the model.
Section 3, which constitutes the main part of this work, presents the analysis based on dynamical mean-field theory (DMFT). In this approach, the retrieval dynamics under interactions are first expressed in terms of generating functions, which are then averaged over random patterns. This procedure reduces the problem to that of a collection of many Langevin equations coupled in a mean-field manner. By solving this ensemble of Langevin equations numerically, we can analyze the behavior of the system for very large system sizes
accurately. 
In Section 4, we compare the results obtained by DMFT with those from direct numerical simulations.
Finally, Section 5 is devoted to summary and discussion.

\section{Associative memory model with non-monotonic transfer functions}
Let us consider a set of $p$ random binary patterns $\bm{\xi}^1, \ldots, \bm{\xi}^p \in \{+1,-1\}^N$ sampled from the uniform distribution over $\{+1,-1\}^{N \times p}$.
In the Hopfield model, these patterns are embedded 
in the synaptic connections among $N$ neurons according to Hebb’s rule as
\begin{align}
J_{ij} = \frac{1}{N}\sum_{\mu=1}^p \xi_i^\mu \xi_j^\mu (1-\delta_{ij}), 
\label{eq:synapse}
\end{align}
\textcolor{black}{where $\delta_{ij}$ denotes the Kronecker delta.} 
Hopfield showed that when an input pattern sufficiently close to one of the stored patterns is presented,
the network dynamics that locally decrease the energy function
\begin{align}
H(\bm{s}) = -\sum_{i<j} J_{ij}s_i s_j,
\label{eq:HopfieldHamiltonian}
\end{align}
where $s_i \in \{+1,-1\}$ represents the state of the $i$-th neuron,
converge to a binary configuration very close to one of the stored patterns.
This convergence corresponds to the retrieval of the stored memory.

A standard retrieval rule is the discrete-time asynchronous update,
\begin{align}
    s_i^{t+1} = \mathrm{sign}\!\left(
    \sum_{j\ne i} J_{ij}s_j^{t}
    \right),
    \label{eq:asynchro}
\end{align}
where $\mathrm{sign}(x)=x/|x|$ for $x\neq 0$.
In this scheme, each neuron updates its state using the \textit{monotonic} transfer function $\mathrm{sign}(a)$,
with input $a_i^t = \sum_{j\ne i} J_{ij}s_j^t$.
{More generally, using a monotonically increasing transfer function $\sigma(x)$ that satisfies
$\sigma(-x) = -\sigma(x)$ and
$\lim_{x \to \pm\infty} \sigma(x) = {\rm sign}(x)$,
the dynamics can be extended to a continuous-time formulation:
\begin{eqnarray}
\label{eq:continuous}
\left\{
\begin{array}{l}
    \displaystyle \frac{d a_i}{d\tau} = -a_i
    + \sum_{j\ne i} J_{ij}\sigma(a_j), \\[6pt]
    s_i = \mathrm{sign}(a_i).
\end{array}
\right.
\end{eqnarray}
This dynamics can also be viewed as a local minimization of the following energy function:
\begin{align}
\label{eq:cont_energy}
    H_{\rm cont}(\bm{a})
    &= -\sum_{i<j} J_{ij}\sigma(a_i)\sigma(a_j)
    + \sum_{i} \int_0^{\sigma(a_i)}
     \sigma^{-1}(v)\,dv,
\end{align}
which is defined for any choice of the 
monotonically increasing transfer function $\sigma(x)$.
Here, $\sigma^{-1}(v)$ denotes the inverse function of $\sigma(x)$.
}

\begin{figure}[t]
\centering
\includegraphics[width=16cm]{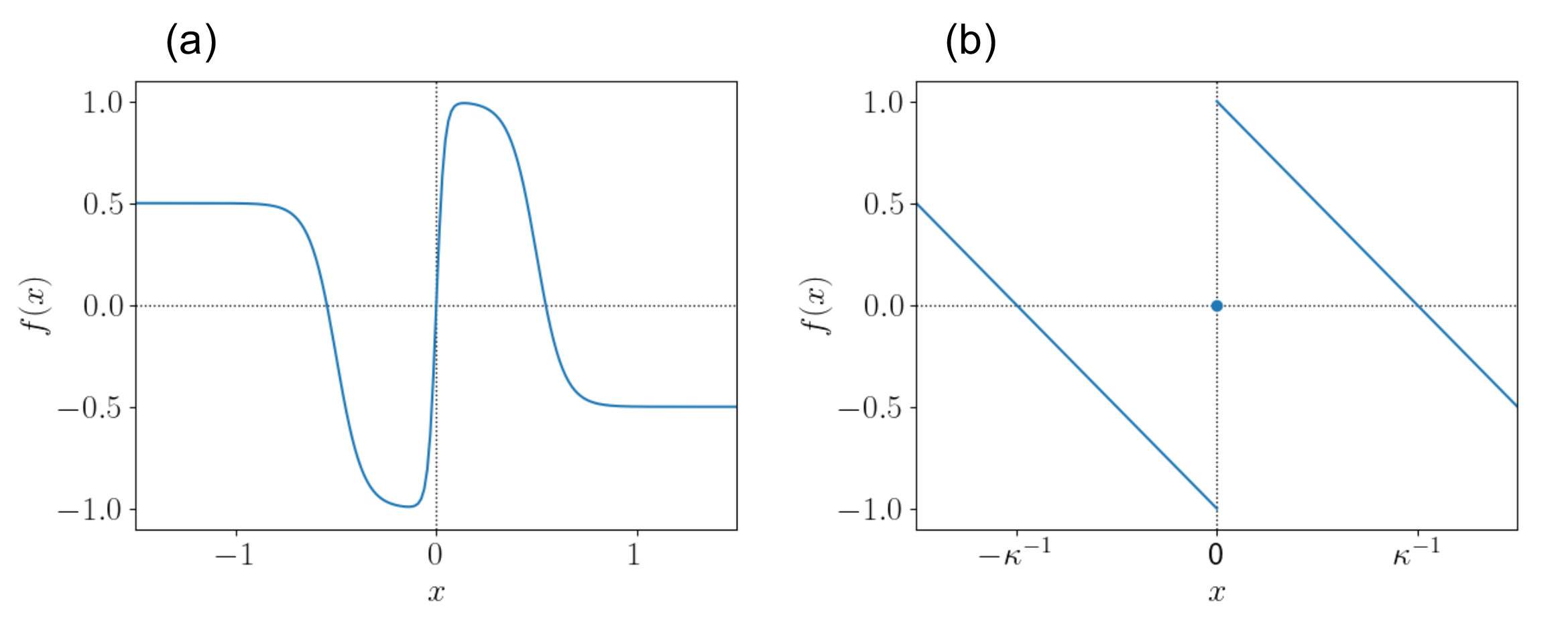}
\caption{\label{fig:fig1} 
(a): Non-monotonic transfer function of Eq. (6). 
(b): Piecewise-linear transfer function. 
}
\end{figure}

These dynamics constitute 
the classical framework of associative memory,
but their retrieval capacity is limited by the gradient-descent nature of the dynamics, which give rise to spurious memories.
Morita (1993) addressed this limitation by replacing 
$\sigma(x)$ in Eq.~(\ref{eq:continuous}) with 
{\em non-monotonic} 
transfer functions $f(x)$~\cite{Morita1993}, such as
\begin{align}
    f(x) = \frac{1-\exp[-cx]}{1+\exp[-cx]}
    \cdot \frac{1+\kappa \exp[c^\prime(|x|-h)]}
    {1+\exp[c^\prime(|x|-h)]}, 
    \label{eq:non-monotonic}
\end{align}
shown in figure \ref{fig:fig1}(a). 
{Because of its non-monotonicity, 
the inverse function $f^{-1}(v)$ cannot 
be defined for Eq.~(\ref{eq:non-monotonic}), 
which makes  
the dynamics no longer described  
by an energy function as Eq.~(\ref{eq:cont_energy}). }
However, the storage capacity is enhanced significantly 
from the conventional value $p_{\rm c} \simeq 0.138N$.
Although its detailed analysis has never been performed, 
he estimated $p_{\rm c}\simeq 0.32N$
based on numerical experiments for $N=1000$. 
He also showed that most spurious states are eliminated, 
yielding qualitatively new retrieval dynamics beyond the conventional energy-based model.
In \cite{Morita1993}, the parameters are tuned around $(c,c^\prime, h, \kappa)=(50,15, 0.5, -1)$, but the resulting retrieval performance 
is robust to these choices. 

The absence of an energy function makes a theoretical analysis of the nonmonotonic model challenging.
Ref. \cite{yoshizawa1993} analyzed a piecewise-linear variant (figure~\ref{fig:fig1}(b)) using a geometrical argument
and showed that the retrieval state remains locally stable and free of spurious states
up to $p_{\rm c} \simeq 0.398N$ under a specific condition. 
Another study~\cite{Brunel1994} defined a cost function that characterizes the fixed point of the retrieval state for a different type of non-monotonic transfer function and, using equilibrium statistical mechanics, demonstrated that the storage capacity increases up to $p_{\rm c} \simeq 0.41N$.
However, these studies are restricted to particular forms of 
$f(x)$ and focus only on local stability,
thus providing no characterization of the full retrieval dynamics or the basin of attraction.
Consequently, a comprehensive theoretical understanding of the Morita model remains an open problem.

\section{DMFT}
For analytical clarity and numerical tractability, we analyze the discrete-time synchronous retrieval dynamics
\begin{eqnarray}
\left \{
\begin{array}{l}
a_i^{t+1} = a_i^t + \gamma
\left [-a_i^t + \sum_{j\ne i} J_{ij} f(a_j^t)\right ], \cr
s_i^t = {\rm sign}(a_i^t),
\end{array}
\right . 
\label{eq:discrete_non_monotonic}
\end{eqnarray}
which reduces to the continuous-time dynamics in the limit
$\gamma\to0$ and $t\to\infty$ while keeping $\tau=\gamma t = O(1)$.
Equation~(\ref{eq:discrete_non_monotonic}) can be expressed by the generating function
\begin{align}
\label{eq:Z}
Z(\tilde{\bm{\theta}}) =
\mathop{\rm tr}_{\bm{s}^0, \tilde{\bm{a}}}
P(\bm{s}^0)
\prod_{t=0}^T \prod_{i=1}^N
\delta\left[a_i^{t+1}-a_i^t-\gamma
\left(-a_i^t+\sum_{j\ne i}J_{ij}f(a_j^t)\right)\right]
\exp\left(i\sum_{t=1}^{T+1}\bm{\theta}^t\cdot\bm{a}^t\right),
\end{align}
which provides the starting point for the dynamical mean-field theory (DMFT) analysis.
Here, at $t=0$, we set $a_i^0 = 0$ 
and replace $f(a_i^0)$ exceptionally 
with the initial state value 
$s_i^0\in \{+1,-1\}$
for $\forall i\in \{1,\ldots,N\}$; 
$P(\bm{s}^0)$ denotes the distribution of $\bm{s}^0 = (s_i^0)$; 
$\tilde{\bm{a}}=(\bm{a}^1,\ldots,\bm{a}^{T+1})$ and
$\tilde{\bm{\theta}}=(\bm{\theta}^1,\ldots,\bm{\theta}^{T+1})$;
and $\mathop{\rm tr}_X(\cdots)$ indicates summation or integration over all configurations of $X$.
The function $Z(\tilde{\bm{\theta}})$ satisfies
\begin{align}
\left. Z(\tilde{\bm{\theta}})\right|_{\tilde{\bm{\theta}}=0}=1, \qquad
\mathbb{E}_{\bm{s}^0, \bm{a}^0}[a_i^t]
=\left.\frac{\partial Z(\tilde{\bm{\theta}})}{\partial (i \theta_i^t)}\right|_{\tilde{\bm{\theta}}=0}, \qquad
\mathbb{E}_{\bm{s}^0, \bm{a}^0}[a_i^t a_j^s]
=\left.\frac{\partial^2 Z(\tilde{\bm{\theta}})}{\partial (i \theta_i^t )\partial (i \theta_j^s)}\right|_{\tilde{\bm{\theta}}=0}, \ldots
\end{align}
where $\mathbb{E}_X[\cdots]$ denotes the expectation with respect to $X$.
Because differentiation and averaging commute, statistical properties of the retrieval dynamics for typical pattern samples
$\tilde{\bm{\xi}}=(\bm{\xi}^1,\ldots,\bm{\xi}^p)$
can be evaluated directly from the averaged generating function
$\mathbb{E}_{\tilde{\bm{\xi}}}[Z(\tilde{\bm{\theta}})]$.

The Fourier representation of the delta function,
\begin{align}
\label{eq:fourier_delta}
&\delta\left[a_i^{t+1}-a_i^t-\gamma
\left(-a_i^t+\sum_{j\ne i}J_{ij}f(a_j^t)\right)\right] \cr
&\quad
=\frac{1}{2\pi i}
\int_{-i\infty}^{+i\infty} dh_i^t
\exp\left [-ih_i^t\left (a_i^{t+1}-a_i^t-\gamma
\left(-a_i^t+\sum_{j\ne i}J_{ij}f(a_j^t)\right)\right)\right],
\end{align}
is the key step for evaluating
$\mathbb{E}_{\tilde{\bm{\xi}}}[Z(\tilde{\bm{\theta}})]$ explicitly.
When the transfer function is antisymmetric,
$f(x)=-f(-x)$, as in Eq.~(\ref{eq:non-monotonic}),
the pattern to be retrieved 
can, without loss of generality, be fixed as
$\bm{\xi}^1 = (1,\ldots,1)^\top$ in analyzing Eq.~(\ref{eq:discrete_non_monotonic}).
Then, the coupling term is decomposed as
\begin{align}
\label{eq:coupling_decomposition}
    \sum_{j\ne i}J_{ij}f(a_j^t)
    &=\frac{1}{N}\xi_i^1\sum_{j=1}^N 
    \xi_j^1f(a_j^t) + \frac{1}{N}\sum_{\mu=2}^p \xi_i^\mu \sum_{j=1}^N \xi_j^\mu f(a_j^t) -\frac{p}{N}f(a_i^t)\cr
    &=m^t + \frac{1}{\sqrt{N}}
    \sum_{\mu=2}^p\xi_i^\mu\sum_{j=1}^N \xi_j^\mu f(a_j^t) -\alpha f(a_i^t), 
\end{align}
where $m^t = N^{-1} \sum_{i=1}^N f(a_i^t)$, $\alpha = p/N$, and the second term in the last 
expression 
$N^{-1/2}\sum_{\mu=2}^p\xi_i^\mu\sum_{j=1}^N \xi_j^\mu f(a_j^t)$
is termed the {\em cross-talk noise}. 
Substituting Eqs.~(\ref{eq:fourier_delta}) and (\ref{eq:coupling_decomposition}) into Eq.~(\ref{eq:Z}) and averaging over $\tilde{\bm{\xi}}$ yield
\begin{align}
\label{eq:mean_Z}
    \mathbb{E}_{\tilde{\bm{\xi}}}
    [Z(\tilde{\bm{\theta}})  ]
    = 
    &\mathop{\rm tr}_{\bm{s}^0, \tilde{\bm{a}}, \tilde{\bm{h}}}
    \left \{
    P(\bm{s }^0)\exp \left [-
    \sum_{t=0}^T\sum_{i=1}^N 
    ih_i^t(a_i^{t+1}-a_i^t-\gamma(-a_i^t  + m^t -\alpha f(a_i^t)) \right ]  \right . \cr
    & \times\left . 
    \mathbb{E}_{\tilde{\bm{\xi}}}
    \left [ \exp \left (\sum_{\mu=2}^{\alpha N} \sum_{t=0}^T \gamma v_\mu^t u_\mu^t \right )
    \right ] \times \exp \left ( i \sum_{t=0}^{T+1}
    \bm{\theta}^t\cdot \bm{a}^t \right )\right \}, 
\end{align}
where 
\begin{align}
v_\mu^t = \frac{1}{\sqrt{N}}\sum_{i=1}^N \xi_i^\mu (ih_i^t),
\qquad
u_\mu^t = \frac{1}{\sqrt{N}}\sum_{j=1}^N \xi_j^\mu f(a_j^t).
\end{align}

For fixed pairs of $\tilde{\bm{a}}$ and $\tilde{\bm{h}}$, the central limit theorem ensures that $u_\mu^t$ and $v_\mu^t$ 
for $\mu \in \{2, \ldots, \alpha N\}$ can be treated as zero-mean multivariate Gaussian variables with covariances
\begin{align}
\label{eq:uu}
    \mathbb{E}_{\tilde{\bm{\xi}}}[u_\mu^t u_\nu^s] 
    = \frac{\delta_{\mu,\nu}}{N}\sum_{i=1}^N f(a_i^t)f(a_i^s)=:\delta_{\mu \nu} Q(t,s),
\end{align}
\begin{align}
\label{eq:vv}
    \mathbb{E}_{\tilde{\bm{\xi}}}[v_\mu^t v_\nu^s] 
    = \frac{\delta_{\mu,\nu}}{N}\sum_{i=1}^N (ih_i^t)(ih_i^s)
    \textcolor{black}{=:} \delta_{\mu\nu} R(t,s),
\end{align}
\begin{align}
    \label{eq:uv}
    \mathbb{E}_{\tilde{\bm{\xi}}}[u_\mu^t v_\nu^s] 
    = \frac{\delta_{\mu,\nu}}{N}\sum_{i=1}^N f(a_i^t)(ih_i^s)=:\delta_{\mu\nu} S(t,s)
\end{align}
for $N\gg 1$, which we assume from now on. 
Assessing 
$\mathbb{E}_{\tilde{\bm{\xi}}}
[ 
\cdots ]$ 
on the right-hand side of 
Eq. (\ref{eq:effective_Z}) by averaging 
over the multivariate Gaussian variables 
and employing the saddle-point method yields
\begin{align}
\label{eq:effective_Z}
   & \mathbb{\rm E}_{\tilde{\bm{\xi}}}
    [Z(\tilde{\bm{\theta}})  ] \cr
    &\simeq 
    \mathop{\rm tr}_{\bm{s}^0, \tilde{\bm{a}}, \tilde{\bm{h}}, \tilde{\bm{\phi}}}
    \left \{
    P(\bm{s}^0)\exp \left [-
    \sum_{t=0}^T\sum_{i=1}^N 
    ih_i^t\left (a_i^{t+1}-a_i^t-\gamma
    \left (-a_i^t + m^t +\phi_i^t +\sum_{s=0}^{t-1}\Lambda(t,s)f(a_i^s) \right )\right ) \right ]  \right . \cr
    & \quad \left . \times 
    \frac{1}{(2\pi)^{\textcolor{black}{N(T+1)/2}}\det(C)^{N/2}}
    \exp\left (-\frac{1}{2}\sum_{i=1}^N
    \sum_{t,s} C^{-1}(t,s) \phi_i^t\phi_i^s
    \right )
    \times 
    \exp \left ( i \sum_{t=0}^{T+1}
    \bm{\theta}^t\cdot \bm{a}^t \right )\right \} \cr
   &= \mathop{\rm tr}_{\bm{s}^0, \tilde{\bm{a}}, \tilde{\bm{\phi}}}
    \left \{
    P(\bm{s}^0)
    \prod_{t=0}^T \prod_{i=1}^N 
    \delta \left [a_i^{t+1}-a_i^t-\gamma
    \left (- a_i^t + m^t +\phi_i^t +\sum_{s=0}^{t-1}\Lambda(t,s)f(a_i^s) \right ) \right ] \right . \cr
    & \quad  \left . \times 
    \frac{1}{(2\pi)^{N(T+1)/2}\det(C)^{N/2}}
    \exp\left (-\frac{1}{2}\sum_{i=1}^N
    \sum_{t,s} C^{-1}(t,s) \phi_i^t\phi_i^s
    \right )
    \times 
    \exp \left ( i \sum_{t=0}^{T+1}
    \bm{\theta}^t\cdot \bm{a}^t \right )\right \} 
\end{align}
where $\tilde{\bm{\phi}} = \{\phi_i^t\}$. 
{This result can be interpreted as the generating function for a set of $N$ 
discrete-time Langevin equations driven by time-correlated noise $\phi_i^t$ and freedback terms:
\begin{align}
\label{eq:Langevin}
a_i^{t+1} = a_i^{t} +
\gamma \left[-a_i^t + m^t + \phi_i^t
+ \sum_{s=0}^{t-1}\Lambda(t,s)f(a_i^s)\right],
\end{align}
where $\mathbb{E}_{\tilde{\bm{\phi}}}[\phi_i^t] = 0$ and
\begin{align}
\label{eq:C2}
\mathbb{E}_{\tilde{\bm{\phi}}}[\phi_i^t \phi_j^s] = \delta_{ij} C(t,s).
\end{align}
Here, $\phi_i^t$ is regarded as the pure noise portion of the cross-talk noise. 
The feedback coefficients $\Lambda(t,s)$ 
physically 
encode the self-feedback effect mediated by the cross-talk noises and the last term of Eq. (\ref{eq:coupling_decomposition}); this mechanism corresponds to Onsager’s reaction term in equilibrium theory~\cite{Thouless1977,Opper2016}.
}

{
The matrices $C=\left (C(t,s) \right )$ and $\Lambda =\left (\Lambda(t,s) \right )$ are
evaluated as follows. }
For a matrix $A$, let $A_t$ denote its
submatrix composed of elements whose row and column indices are less than or equal to $t$. 
Using this notation, the matrix $C $ is constructed by recursively adding one row and one column at each time step, as 
\begin{align}
   & C_t(t,s) =C_t(s,t)= \alpha \!\left[(I_t - G_t)^{-1} Q_t (I_t - G_t^\top)^{-1}\right]\!(t,s), \label{eq:C}\\
   & G_t(t,s) =
   \begin{cases}
   \displaystyle
   \sum_{w < t} \!\left(\frac{1}{N} \sum_{i=1}^N f(a_i^t)\phi_i^w \right)\! C_{t-1}^{-1}(w,s), & s < t, \\[8pt]
   0, & s=t,
   \end{cases} \label{eq:G}
\end{align}
with initial conditions $C(0,0) = \alpha$ and $G(0,0) = 0$, 
where $I_t$ denotes the identity matrix of size $t\times t$. 
{Quantities 
$G(t,s)$ are referred to as the 
response functions since they coincide with  
$N^{-1} \sum_{i=1}^N \mathbb{E_{\tilde{\bm{\phi}}}} \left [
\partial f(a_i^t)/\partial \phi_i^s \right ]$
for $N\to \infty $ (see Eq.~(\ref{eq:Response_derivation})). 
Due to the nature of the construction method, the matrix $G=(G(t,s))$ becomes a lower-triangular matrix. }
The feedback coefficients are then evaluated recursively as 
\begin{align}
\Lambda_t(t,s) =
\begin{cases}
\alpha \!\left[G_t (I_t - G_t)^{-1}\right]\!(t,s), & s < t, \\[6pt]
0, & \text{otherwise}.
\end{cases}
\label{eq:Gamma}
\end{align}
The derivation of Eqs. (\ref{eq:effective_Z})--(\ref{eq:Gamma}), 
{which constitute the main 
result of the present paper}, 
is provided in Appendix~\ref{appendix1}.


{Eqs. (\ref{eq:Langevin})--(\ref{eq:Gamma})
}
involve no direct interactions among neurons.
Consequently, numerical analysis 
{based on these equations}
can be performed for much larger system sizes $N$ than would be feasible for direct simulations of Eq.~(\ref{eq:discrete_non_monotonic}).
Since the saddle-point method used in deriving Eq.~(\ref{eq:effective_Z}) yields exact results in the limit $N \to \infty$,
the remaining numerical errors in Eq.~(\ref{eq:Langevin}) originate solely from the sampling of the random process $\tilde{\bm{\phi}}$~\cite{Eissfeller1992}.
This feature makes the present framework highly suitable for accurately investigating the macroscopic behavior of large systems.

On the other hand, evaluating $C(t,s)$ and $\Lambda(t,s)$ requires matrix inversion at each time step,
which scales computationally as \textcolor{black}{$O(t^3)$} up to the $t$-th time step \textcolor{black}{even when tricks based on the Shur complement formula are employed~\cite{Zhang2006}.}
This makes it practically difficult to carry out numerical analyses for $t \gtrsim O(10^3)$ even with modern computational resources.
Accordingly, it remains challenging to directly apply the present method to the continuous-time limit,
where $t \to \infty$ and $\gamma \to 0$ are taken while keeping $\tau = \gamma t = O(1)$.

\section{Results}
\subsection{Comparison between DMFT and direct Simulations}
\label{sec:comparison}
We compared the results obtained from the DMFT effective dynamics Eq.~(\ref{eq:Langevin}) with those from direct simulations of Eq.~(\ref{eq:discrete_non_monotonic}). 
For the neuronal transfer function, we used Eq.~(\ref{eq:non-monotonic}) with the parameter set 
$(c, c', h, k) = (50, 15, 0.5, -0.5)$, as specified in Morita’s original work~\cite{Morita1993}. 
The system sizes were set to $N_{\rm DMFT} = 10^6$ and $N_{\rm direct} = 4096$ for the DMFT and direct simulations, respectively, and the time step size was fixed at $\gamma = 0.1$. 

\begin{figure}[t]
\centering
\includegraphics[width=16cm]{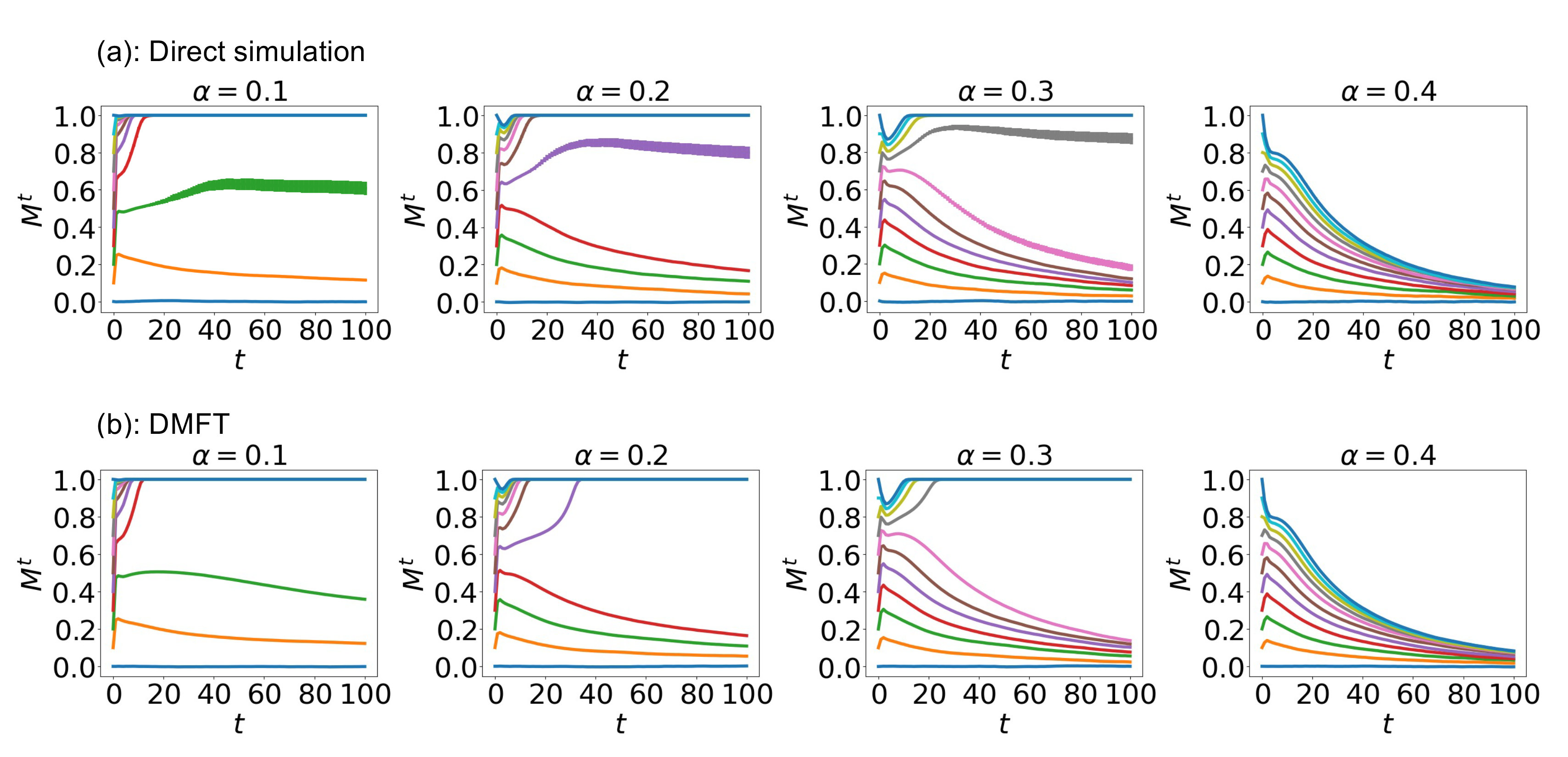}
\caption{\label{fig:fig2} Comparison of the readout overlap $M^t$ between (a) direct simulations and (b) DMFT results.}
\end{figure}

\begin{figure}[t]
\centering
\includegraphics[width=16cm]{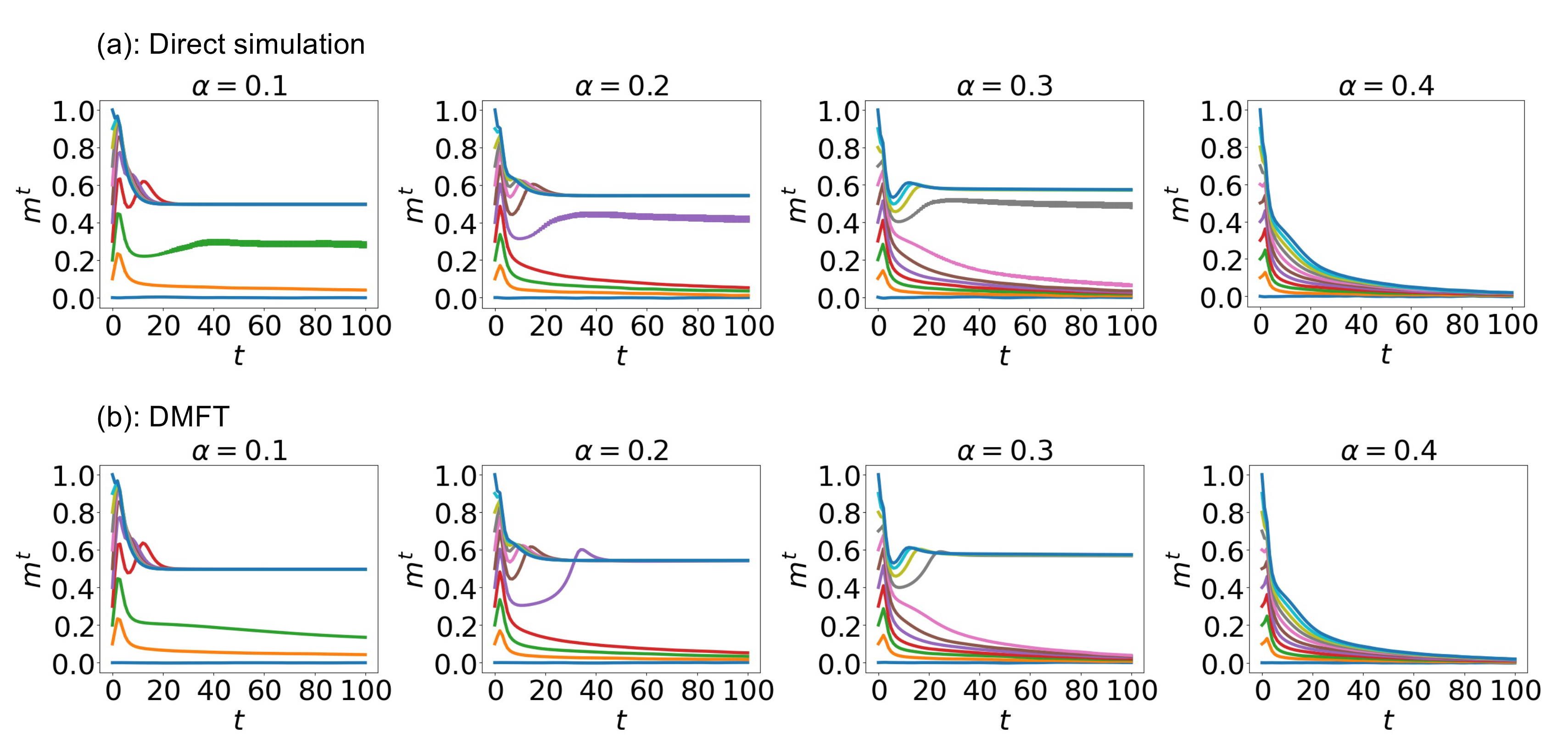}
\caption{\label{fig:fig3} Comparison of the output overlap $m^t$ between (a) direct simulations and (b) DMFT results.}
\end{figure}

Figure~\ref{fig:fig2} compares the readout overlap 
\[
M^t = N^{-1} \sum_{i=1}^N \xi_i^1 s_i^t = N^{-1} \sum_{i=1}^N {\rm sign}(a_i^t)
\]
between (a) the direct simulations and (b) the DMFT results. 
A similar comparison for the output overlap, 
\[
m^t = N^{-1} \sum_{i=1}^N \xi_i^1 f(a_i^t) = N^{-1} \sum_{i=1}^N f(a_i^t),
\]
is shown in figure~\ref{fig:fig3}. 
Except for the critical cases, excellent agreement is observed in both figures, confirming that the DMFT approach reliably captures the macroscopic dynamical behavior of large systems. 

Figure~\ref{fig:fig4} shows the readout overlap $M^t$ at $t = 100$ as a heat map. 
The horizontal and vertical axes represent the pattern ratio $\alpha$ and the initial overlap $M^0$, respectively. 
Since the success or failure of retrieval can be 
almost determined by $t = 100$, this figure can be regarded as an empirical basin of attraction. 
It indicates that the storage capacity reaches $\alpha_{\rm c} = p_{\rm c}/N \simeq 0.36$, 
slightly higher than Morita’s estimate of $0.32$ \cite{Morita1993}. 
This difference is likely due to finite-size effects, 
as Morita’s result was obtained from numerical experiments on relatively small systems ($N = 1000$). 

{
The expansion of the basin of attraction induced by non-monotonic transfer functions can be understood as follows~\cite{yoshizawa1993}.
Increasing the number of stored patterns enlarges the variance of the cross-talk noise. As a result, 
 an increasing number of neurons receive large inputs with the opposite sign from that of the target pattern to be retrieved. 
 When a monotonic transfer function is used,
 the presence of many such neurons that are “overconfident” yet incorrect undermines the stability of the retrieval state.
In contrast, with a non-monotonic transfer function, neurons receiving excessively large inputs flip their output. Consequently, such overconfident neurons cannot persist, which makes the variance of the cross-talk noise remain relatively small compared with the signal components that convey information about the target pattern. This allows the retrieval state to remain stable up to larger
pattern ratios $\alpha$ 
than in the case with a monotonic transfer function.
}

\subsection{Assessment of self-feedback effects}
\begin{figure}[t]
\centering
\includegraphics[width=16cm]{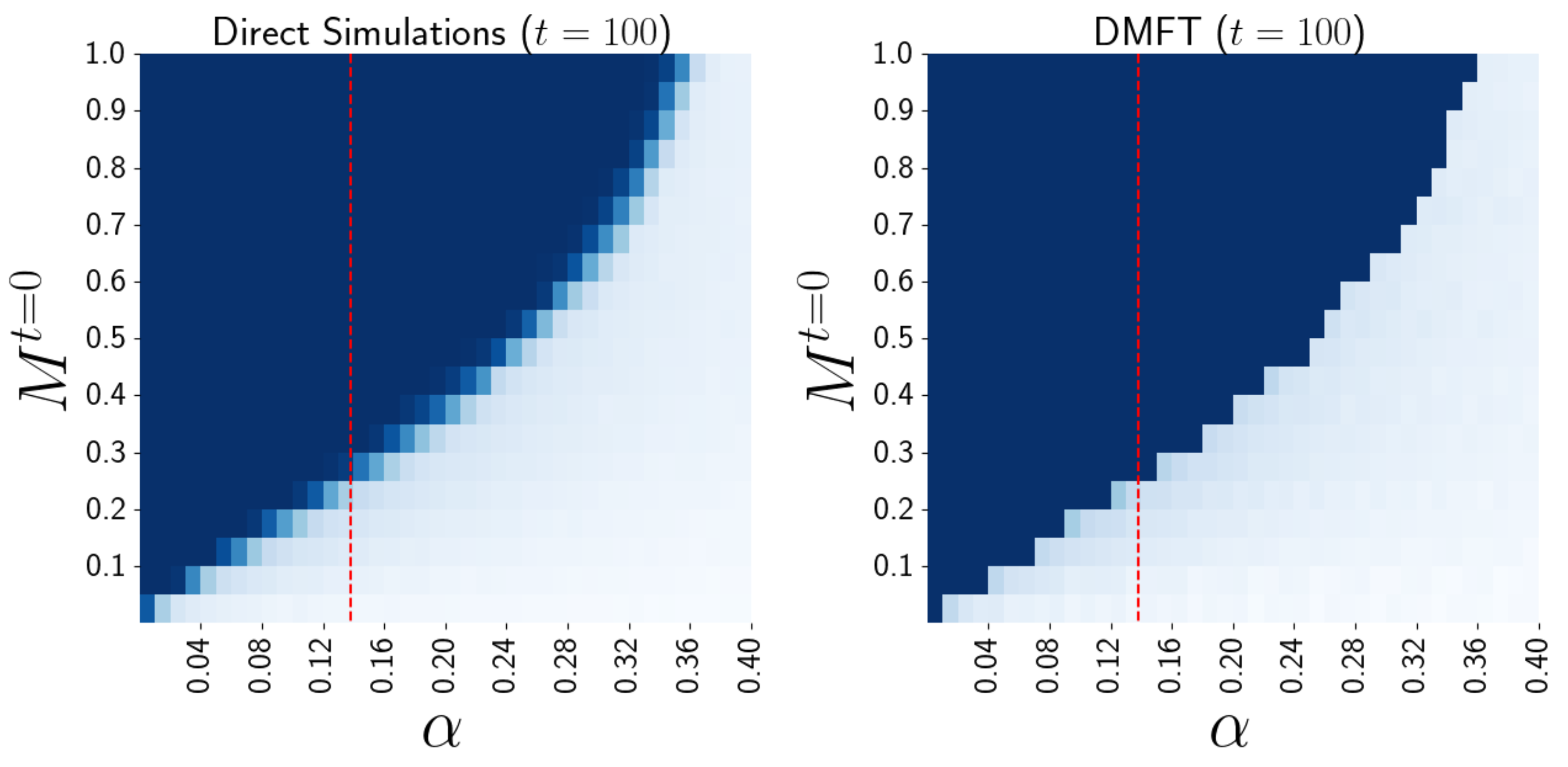}
\caption{\label{fig:fig4} Heat map of the readout overlap $M^t$ at $t = 100$. 
The horizontal and vertical axes correspond to $\alpha$ and $M^0$, respectively. Vertical broken lines stand for the conventional storage capacity 
$\alpha_{\rm c} \simeq 0.138$.}
\end{figure}

\begin{figure}[htbp]
\centering
\includegraphics[width=16cm]{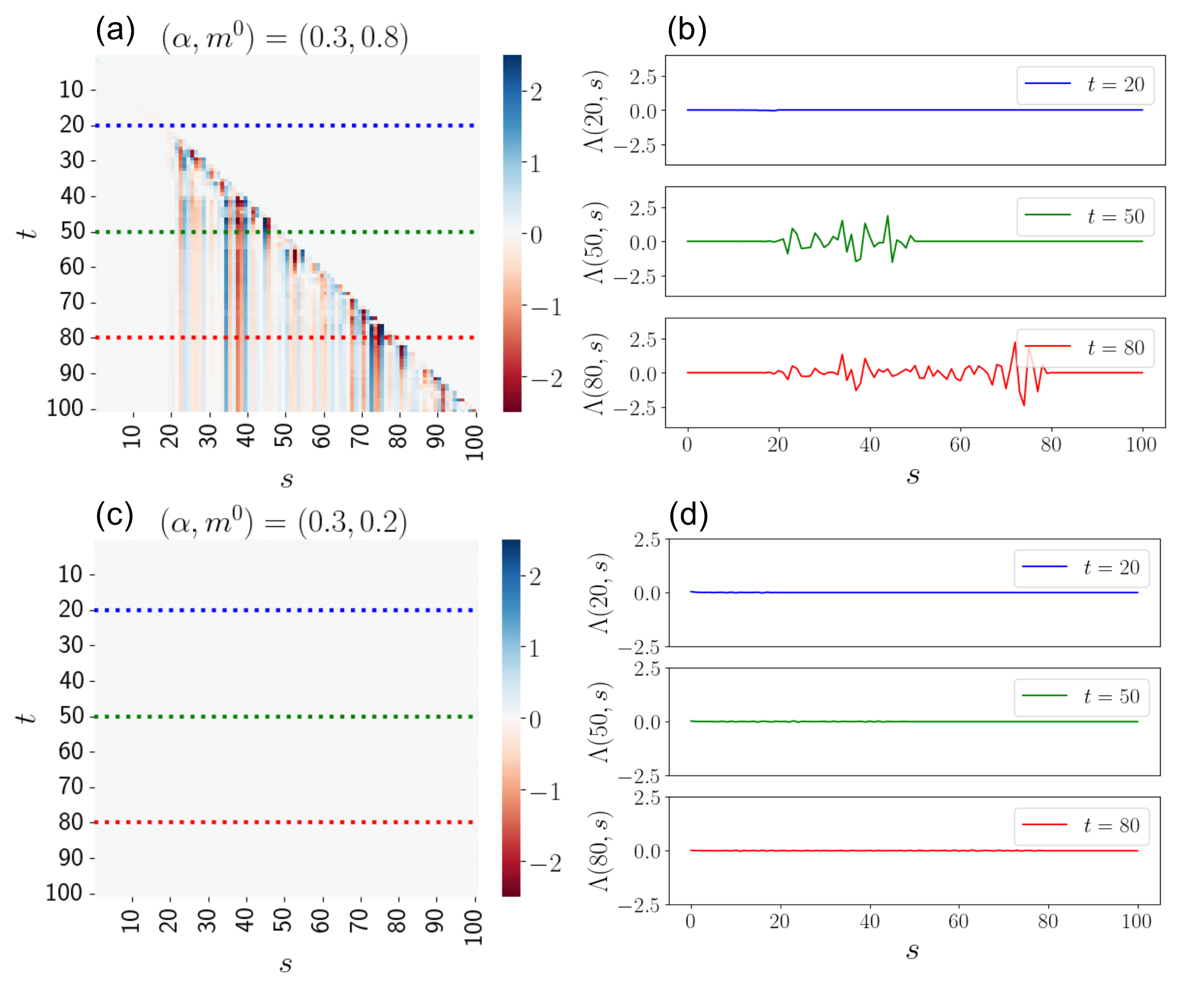}
\caption{\label{fig:fig5} Time evolution of the feedback coefficients $\Lambda(t,s)$. (a,b): Successful retrieval. (c,d): Failed retrieval.
Broken lines in (a) and (c) 
represent positions of data plotted 
in (b) and (d). 
To facilitate comparison, identical plot scales were used for both the successful and failed retrieval cases.
}
\end{figure}

Beyond enabling access to the macroscopic properties of large systems, 
another advantage of the DMFT framework is its ability to evaluate self-feedback effects explicitly. 
Figure~\ref{fig:fig5} plots the feedback coefficients $\Lambda(t,s)$ for both successful (panels (a), (b)) and failed (panels (c), (d)) retrieval cases. 
When retrieval succeeds, $\Lambda(t,s)$ takes non-negligible values once the system has macroscopically converged to the retrieval state (around $t \gtrsim 20$, see figures~\ref{fig:fig3} and~\ref{fig:fig4}) and continues to exhibit irregular oscillations thereafter. 
Although the retrieval state appears macroscopically stationary and seems to satisfy time-translation invariance, $\Lambda(t,s)$ itself is not stationary. 

\begin{figure}[t]
\centering
\includegraphics[width=16cm]{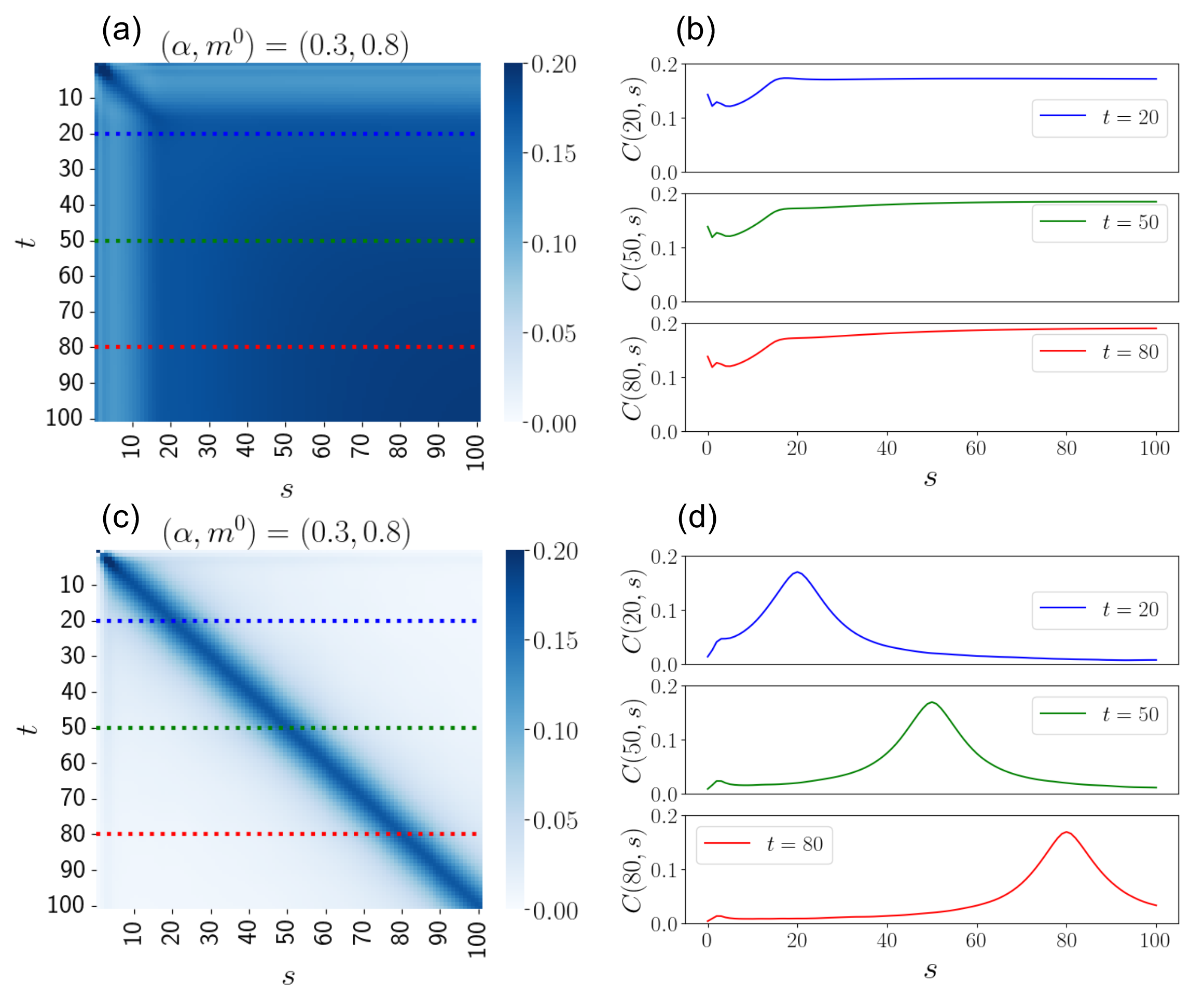}
\caption{\label{fig:fig6} Time correlation matrix of noises $C(t,s)$ during retrieval.
(a,b): Successful retrieval. (c,d): Failed retrieval.
Broken lines in (a) and (c) 
represent positions of data plotted 
in (b) and (d). 
To facilitate comparison, identical plot scales were used for both the successful and failed retrieval cases.
}
\end{figure}

This apparent irregularity is, however, not intrinsic to the dynamics itself but can be attributed to Monte Carlo errors arising from the stochastic nature of the Langevin updates.
When the system converges to a retrieval state, the random variable $\phi_i^t$—originating from the cross-talk noise
$N^{-1/2} \sum_{\mu=2}^p \xi_i^\mu \sum_{j=1}^N \xi_j^\mu f(a_j^t)$—also converges to a constant.
Consequently, the noise correlation approaches a finite value in the long-time limit (figures \ref{fig:fig6} (a,b)), 
$C(t,s) \to \sigma_{\phi}^2$ as $t,s \to \infty$, 
{and 
the correlation between $f(a_i^t)$ and $\phi_i^t$,
}
$K(t,s) = N^{-1} \sum_{i=1}^N 
f(a_i^t)\phi_i^s$, 
{
which appears in the right-hand side
in Eq.~(\ref{eq:G}), also}
tends to a constant $K^*$ in the same limit.
This implies that, for large $t$, the matrix $C_{t-1}$ possesses very small eigenvalues
in directions orthogonal to $(1,\ldots,1)^\top$.
As a result, when $C_{t-1}^{-1}$ acts in Eq.~(\ref{eq:G}),
Monte Carlo fluctuations along those orthogonal directions in $G_t(t,s)$ are strongly amplified,
which in turn leads to large fluctuations in $\Lambda_t(t,s)$ through Eq.~(\ref{eq:Gamma}).

For the asymptotic regime where $t, s \gg 1$, the component of $\Lambda(t,s)$ (which is equal to $\Lambda_t(t,s)$) with the fluctuation term removed is evaluated as follows. In this context, it is reasonable to assume $C(t,s) = \sigma_\phi^2$ and $K(t,s) = K^*$
for all pairs of $t$ and $s$. 
The component of $G(t,s)$ in the direction of $(1,\ldots,1)^\top$ is then derived from Eq.~(\ref{eq:G}) as:
\begin{align}
\label{eq:G_t_limit}
G(t,s) =
\begin{cases}
\displaystyle
 \frac{ \sigma_{\phi}^{-2}K^*}{t }, & s< t, \\
0, & s\ge t.
\end{cases}
\end{align}
Consequently, 
$\Lambda(t,s) =
\alpha \left [G_t (I_t-G_t)^{-1} \right ](t,s)$ is given by:
\begin{align}
\label{eq:Lambda_ts_asympt}
\Lambda(t,s) &= \Lambda_t(t,s)\cr
&= 
\begin{cases}
\displaystyle
\frac{\alpha \sigma_\phi^{-2} K^*}{t}
\frac{\Gamma(t+ \sigma_\phi^{-2} K^*)}{\Gamma(t)}
\frac{\Gamma(s+1)}{\Gamma(s+1 + \sigma_\phi^{-2} K^*)}, &s< t, \cr
0, & s\ge t
\end{cases}
\cr
&\simeq 
\begin{cases}
\displaystyle
\frac{\alpha \sigma_\phi^{-2} K^*}{t}
\left (
\frac{t}{s}
\right )^{\sigma_\phi^{-2} K^*}, &
s< t,\cr
0, & s\ge t, 
\end{cases}
\end{align}
where $\Gamma(x)$ denotes the Gamma function.
See Appendix \ref{appendixLambda} for the derivation.

\begin{figure}[t]
\centering
\includegraphics[width=16cm]{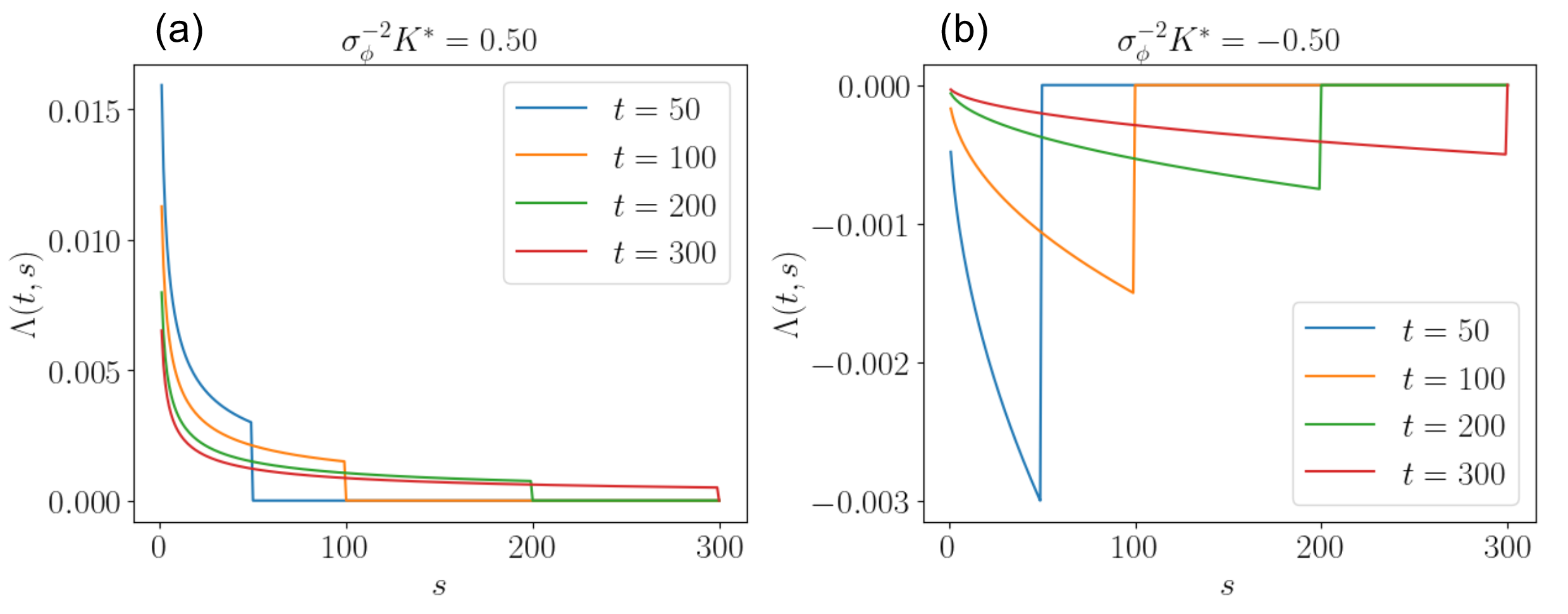}
\caption{\label{fig:fig7} Example profiles of Eq. (\ref{eq:Lambda_ts_asympt})
for (a): $K^*>0$ and (b): $K^* < 0$. 
The pattern ratio $\alpha$ is set $0.3$ in both 
cases. 
}
\end{figure}

The behavior of Eq.~(\ref{eq:Lambda_ts_asympt}) 
depends on the sign of $K^*$. 
When $K^*>0$ the feedback coeffcient $\Lambda(t,s)$ of 
past states at time $s<t$
on a given time $t$ becomes stronger as one goes further back in time.
In contrast, when $K^*<0$, the most recent past exerts the strongest effect, and the influence gradually decays as one traces further back in time.
However, in either case, 
individual feedback components from past times 
$s$ vanish algebraically as $t$ increases
(figures \ref{fig:fig7} (a,b)).

\begin{figure}[t]
\centering
\includegraphics[width=16cm]{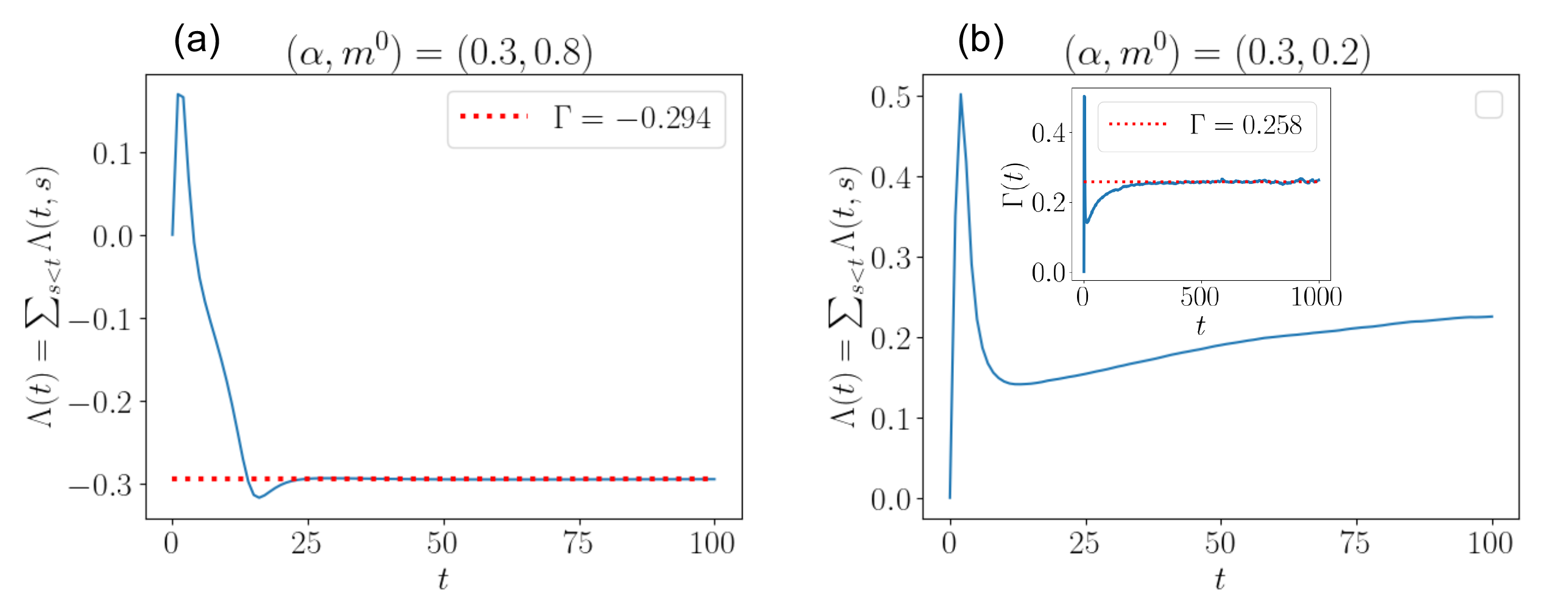}
\caption{\label{fig:fig8} Temporal evolution of the integrated feedback coefficient $\Lambda(t)$. 
(a): Successful retrieval. (b): Failed retrieval.}
\end{figure}

Unfortunately, 
because of large fluctuations, it is practically challenging to directly verify the validity of Eq.~(\ref{eq:Lambda_ts_asympt}).
Nevertheless, Eq.~(\ref{eq:Lambda_ts_asympt}) suggests that the integrated feedback coefficient is given by
\begin{align}
\label{eq:integrated_Lambda}
\Lambda(t) &= \sum_{s<t} \Lambda(t,s) \cr
&\simeq\frac{\alpha \sigma_\phi^{-2} K^*}{t}
t^{\sigma_\phi^{-2}K^*}
\sum_{s<t } s^{-\sigma_\phi^{-2} K^*} \cr
&=\frac{\alpha \sigma_\phi^{-2} K^*}{t}
\times
t^{\sigma_\phi^{-2}K^*}\times
\frac{t^{1-\sigma_\phi^{-2}K^*}-1}{1-\sigma_\phi^{-2}K^*},
\end{align}
and therefore it converges to a constant value
\begin{align}
\label{eq:Lambda_converged}
    \Lambda = \frac{\alpha \sigma_\phi^{-2}K^* }{1-\sigma_\phi^{-2}K^* }
\end{align}
for $t \gg 1$ when $\sigma_\phi^{-2}K^* < 1$, which we assume here.
Furthermore, since the fluctuations are orthogonal to $(1,\ldots,1)^\top$, they are expected to cancel out in Eq. (\ref{eq:integrated_Lambda}).
Indeed, the behavior of $\Lambda(t)$ observed in successful retrieval cases supports this scenario (figure \ref{fig:fig8}~(a)).

\begin{figure}
    \centering
    \includegraphics[width=14cm]{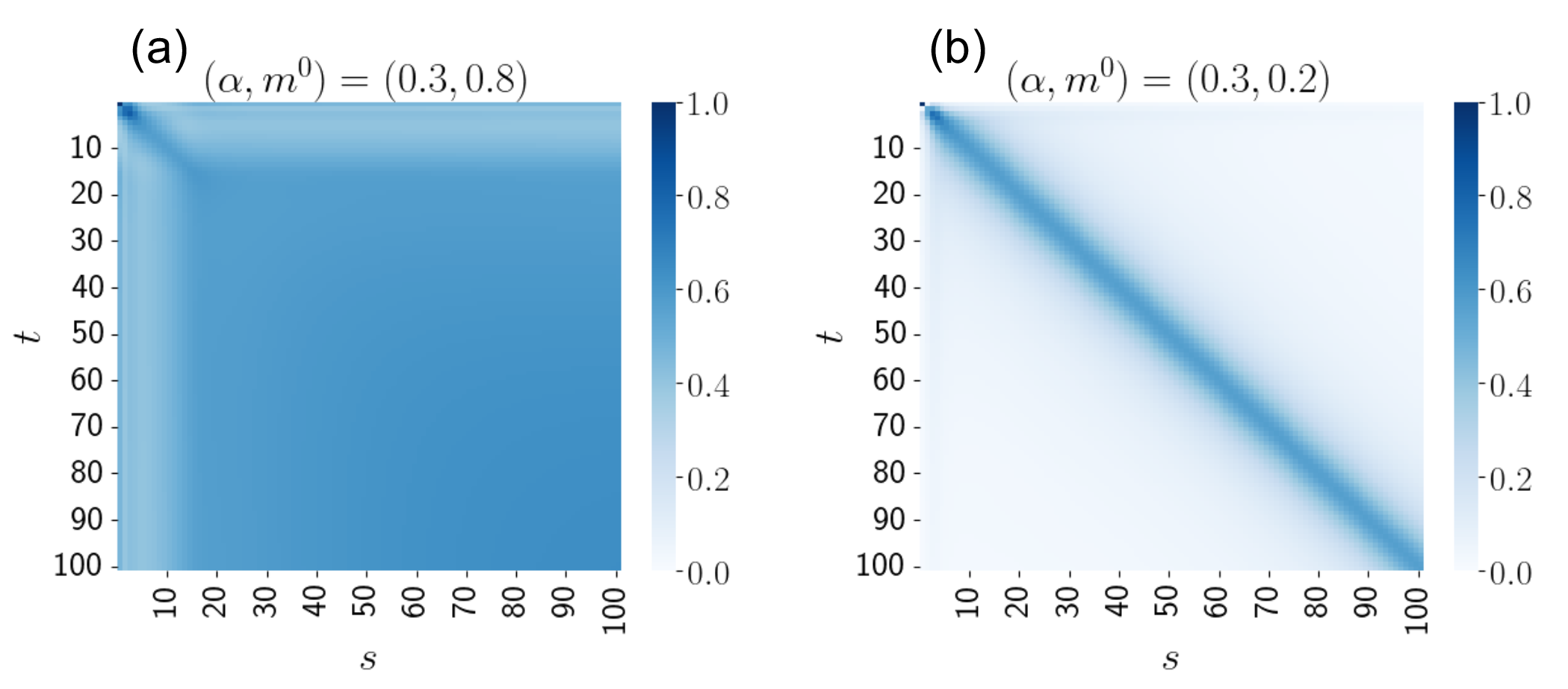}
    \caption{
    {Time correlation 
    \label{fig:fig9}
    $Q(t,s) = N_{\rm DMFT}^{-1}  \sum_{i=1}^Nf(a_i^t)f(a_i^s)$.
    (a): Successful retrieval. 
    (b): Failed retrieval. }
    }
\end{figure}

{
Conversely, when retrieval fails, the correlation function $C(t,s)$ decays as 
$|t-s|$ increases, exhibiting stationary behavior 
$C(t,s) = C(|t-s|)$
(figures~\ref{fig:fig6} (c,d)).
The feedback coefficients $\Lambda(t,s)$ remain small except for the first few steps of $t$, 
while their integrated form continues to evolve slowly 
(figure~\ref{fig:fig8} (b)), eventually converging to a finite value at longer
time scales (inset of figure~\ref{fig:fig8} (b)).
Figure \ref{fig:fig9} shows the time correlation of $f(a_i^t)$, $Q(t,s)= N_{\rm DMFT}^{-1}  \sum_{i=1}^Nf(a_i^t)f(a_i^s)$. 
In contrast to the case of 
successful retrieval (figure~\ref{fig:fig9} (a)), the time correlation $Q(t,s)$ decays rapidly as $|t-s|$ grows
in the failed retrieval case (figures~\ref{fig:fig9} (b)). 
These observations suggest that the system stays in a chaotic state.
}

\subsection{Connection to earlier studies}
By employing the cavity method~\cite{Mezard1987}, 
the convergent behavior of $\Lambda(t) = \sum_{s < t} \Lambda(t,s) \to \Lambda$ observed in the successful retrieval cases, together with Eq.~(\ref{eq:Langevin}), 
leads to the following conditions that the fixed point (the retrieval state) must satisfy: 
\begin{align}
    m &= \frac{1}{N_{\rm DMFT}} \sum_{i=1}^{N_{\rm DMFT}} g(m + \phi_i), \label{eq:m_macro}\\
    U &= \frac{1}{N_{\rm DMFT}} \sum_{i=1}^{N_{\rm DMFT}} g'(m + \phi_i), \label{eq:U_macro}\\
    \sigma_\phi^2 &= \frac{\alpha}{N_{\rm DMFT}} \sum_{i=1}^{N_{\rm DMFT}} 
    \frac{[g(m + \phi_i)]^2}{(1 - U)^2}, \label{eq:sig2_macro}\\
    \Lambda &= \frac{\alpha U}{1 - U}. \label{eq:Gamma_macro}
\end{align}
Here, $\phi_i \sim \mathcal{N}(0, \sigma_\phi^2)$, and the function $g(x)$ is determined by solving 
\begin{align}
    g(x) = f(x + \Lambda g(x))
    \label{eq:g_function}
\end{align}
for each $x$.
{
The derivation of Eqs.~(\ref{eq:m_macro})--(\ref{eq:g_function}) is
provided in Appendix \ref{appendix2}.
}
The convergent value of 
the readout overlap is 
assessed as $M = N_{\rm DMFT}^{-1}\sum_{i=1}^{N_{\rm DMFT}}
{\rm sign}(m+\phi_i)$. 
Comparison of Eqs. (\ref{eq:Lambda_converged}) and 
(\ref{eq:Gamma_macro}) indicates that 
$U = \sigma_\phi^{-2}K^*$ holds.
Eqs.~(\ref{eq:m_macro})--(\ref{eq:g_function}) are equivalent to the self-consistent signal-to-noise analysis (SCSNA) developed for analog neural networks in Ref.~\cite{Shiino1993}, 
provided that the sample averages $N_{\rm DMFT}^{-1} \sum_{i=1}^{N_{\rm DMFT}} (\cdots)$ 
are replaced by Gaussian averages 
$\int d\phi\, \mathcal{N}(\phi|0, \sigma_\phi^2)(\cdots)$, which is justified by the law of large numbers in the limit of $N_{\rm DMFT} \to \infty$.  
Similar results can also be found in \cite{Waugh1993}.

\begin{figure}[t]
\centering
\includegraphics[width=14cm]{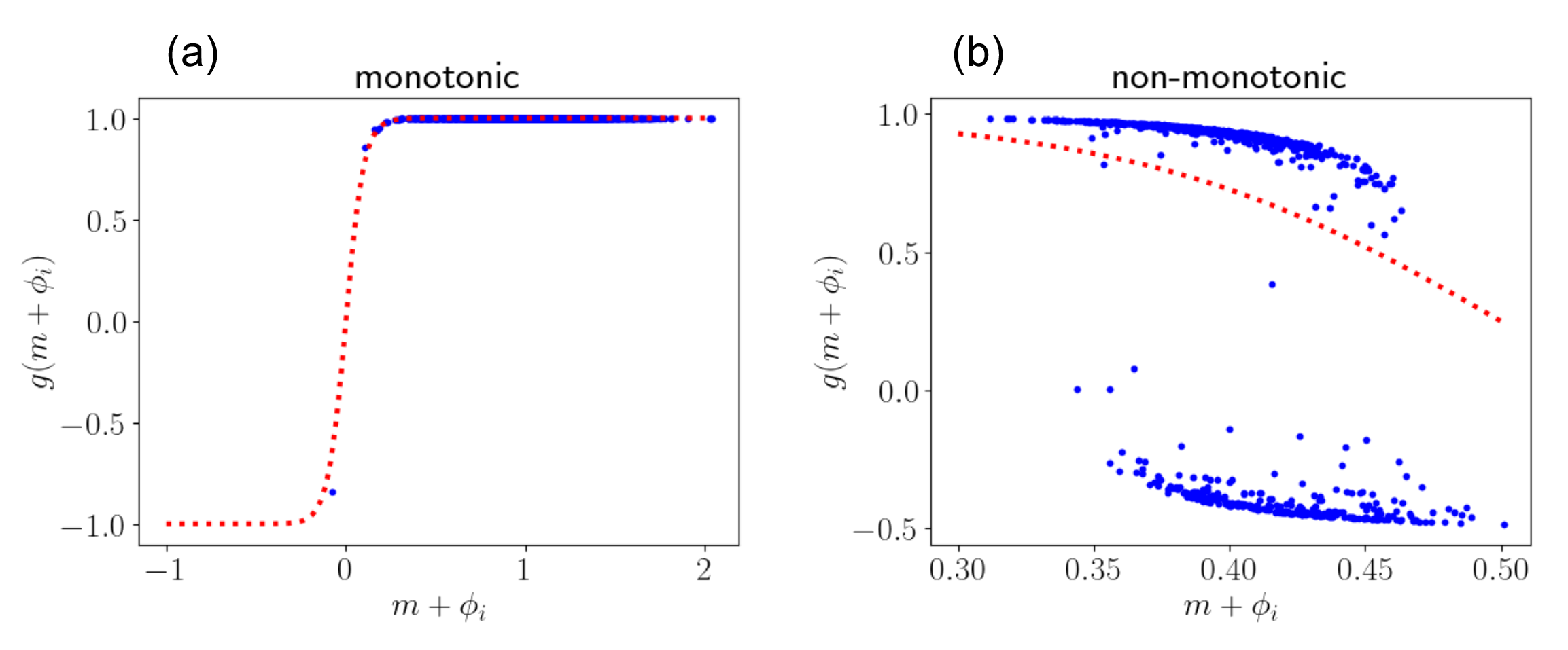}
\caption{
{\label{fig:fig10} 
Markers show input-output relations after 1000 iterations of Eq.~(\ref{eq:Langevin}), which practically
correspond to the solutions $g(x)$ of Eq.~(\ref{eq:g_function}).
(a) Monotonic transfer function $\tanh(10x)$ with $\alpha = 0.1$.
(b) Non-monotonic transfer function given by Eq.~(\ref{eq:non-monotonic}) with $\alpha = 0.3$.
In both cases, the initial state was set to the target pattern $(1, 1, \ldots, 1)^\top$.
In contrast to the monotonic case (a), multiple branches of $g(x)$ emerge in the non-monotonic case (b). Red dotted lines stand for the transfer functions. 
}}
\end{figure}

Ref.~\cite{Shiino1993} shows that SCSNA can accurately describe macroscopic quantities such as storage capacity and the local-field distribution 
as long as the function $g(x)$ is uniquely determined by Eq.~(\ref{eq:g_function}). 
However, it is also reported that when multiple solutions of Eq.~(\ref{eq:g_function}) emerge as the profile of $f(x)$ is gradually modified, 
the SCSNA predictions become unreliable. 
Indeed, in the present model, the convergent DMFT solution reveals that Eq.~(\ref{eq:g_function}) admits multiple branches {while $g(x)$ is a single valued function for a monotonic transfer function (figures~\ref{fig:fig10} (a,b))}
For each input $m + \phi_i$ in Eqs.~(\ref{eq:m_macro})--(\ref{eq:sig2_macro}), 
the appropriate branch of $g(m + \phi_i)$ is determined by the entire history of the DMFT dynamics. 
This implies that DMFT is indispensable for examining the fixed-point structure of the retrieval dynamics
{when non-monotonic transfer functions are used.}

\section{Summary and Discussion}

In this study, we applied the dynamical mean-field theory (DMFT) to the associative memory model with non-monotonic transfer functions originally proposed by Morita~\cite{Morita1993}.  
This model, despite its remarkably high retrieval performance, lacks an energy function and thus cannot be analyzed within the framework of equilibrium statistical mechanics.  
By reformulating the dynamics as a collection of Langevin equations with time-correlated noise and self-consistent feedback,  
we succeeded in quantitatively characterizing its macroscopic retrieval dynamics.

The comparison between DMFT and direct numerical simulations demonstrated excellent agreement,  
establishing the validity of the DMFT approach even for associative memory models that are not characterized by energy functions.  
For a discrete-time synchronous update with a time step $\gamma = 0.1$,  
the storage capacity was found to reach $\alpha_{\rm c} \simeq 0.36$,  
which is slightly larger than the original estimate by Morita ($\alpha_{\rm c} \simeq 0.32$).  
This discrepancy can be attributed to finite-size effects in the earlier simulations.

\textcolor{black}{A distinctive aspect revealed by our analysis lies in the behavior of the feedback coefficient 
in the retrieval state.
In the DMFT framework, macroscopic quantities are evaluated through a large number of Langevin equations, making it impossible to completely eliminate Monte Carlo errors.
Since the feedback coefficient is highly sensitive to such statistical fluctuations, obtaining precise numerical estimates is extremely challenging.
However, after theoretically eliminating the influence of these fluctuations, we were able to quantify the 
asymptotic form of the feedback from past states on the system at a given observation time in the retrieval state.
}

Based on the behavior of the feedback coefficient, we also derived the fixed-point conditions satisfied by the retrieval dynamics, which reduce to those obtained from the self-consistent signal-to-noise analysis (SCSNA) \cite{Shiino1993} when the microscopic averages are replaced by Gaussian averages. 
However, we found that in the present model the function $g(x)$, which is determined by a functional equation $g(x) = f(x + \Lambda g(x))$, develops multiple branches,  
and the appropriate branch depends on the dynamical history of the system.  
This observation explains why SCSNA fails to capture the full structure of the retrieval fixed points,  
whereas DMFT remains valid even in such nontrivial regimes.

The present results demonstrate that DMFT provides a powerful framework for studying non-equilibrium neural dynamics that cannot be analyzed within traditional energy-based approaches.  
Extending this framework 
to models with correlated patterns  
and to learning systems with structured or time-dependent synaptic modifications  
represents a promising direction for future work.  
Such extensions may further bridge the gap between statistical-physics-based memory models and modern theories of learning and inference in complex neural systems.
Another interesting direction is to revisit the two-parameter dynamical replica theory (2-DRT) of Tanaka, Kakiya, and Okada \cite{Tanaka2001}, which successfully described retrieval dynamics and super-retrieval states in a non-monotonic associative memory model. Comparing its predictions with those of DMFT, which does not require the equipartitioning assumption underlying 2-DRT, may provide further insights into the mechanisms of ``super-retrieval'', for which patterns are readout completely without any errors, and the validity of approximate macroscopic descriptions of non-equilibrium dynamics.

\begin{acknowledgments}
This work was partially supported by 
MEXT/JSPS KAKENHI Grant No. 22H05117 (YK) 
and JSPS KAKENHI Grant Nos. 23H05492 and 23K03841 (KM). 
\end{acknowledgments}

\appendix 
\section{Derivation of Eqs. (\ref{eq:effective_Z})--(\ref{eq:Gamma})}
\label{appendix1}
Let us denote $(T+1)\times (T+1)$ 
matrices composed of 
$Q(t,s)$, $R(t,s)$, and $S(t,s)$ in 
Eqs. (\ref{eq:uu})--(\ref{eq:uv}) by
$Q$, $R$, and $S$, respectively.
Using these quantities, 
we define a $2(T+1)\times 2(T+1)$ matrix as
\[
\Sigma =
\left (
    \begin{array}{cc}
    Q & S \cr
    S^\top & R
    \end{array}
\right ).
\]

Evaluating $
\mathbb{E}_{\tilde{\bm{\xi}}}
    \left [ \cdots
    \right ]$
on the right-hand side of Eq. (\ref{eq:mean_Z}) 
by averaging over multivariate Gaussian variables
whose covariances are given by 
Eqs. (\ref{eq:uu})-(\ref{eq:uv}) yields
\begin{small}
\begin{align}
\label{eq:exp_gamma_uv}
&\mathbb{E}_{\tilde{\bm{\xi}}}\!
 \left[\exp\!\left(\sum_{\mu=2}^{\alpha N}\sum_{t=0}^T 
 \gamma v_\mu^t u_\mu^t\right)\right]\cr
&=\!\left[
\frac{1}{(2\pi)^{(T+1)}(\det\Sigma)^{1/2}}
\!\int\!
\exp\!\left(
-\tfrac12(\bm{u}^\top \ \bm{v}^\top)
\Sigma^{-1}\!\!
\begin{pmatrix}\bm{u}\\\bm{v}\end{pmatrix}
+\gamma\bm{v}^\top\bm{u}
\right)\!
\prod_{t=0}^T\!du^t dv^t
\right]^{\!\alpha N}\cr
&=\!\left[
\frac{1}{(2\pi)^{2(T+1)}}\!\!
\int\!
\exp\!\left(
-\tfrac12(\bm{x}^\top \ \bm{y}^\top)
\Sigma\!
\begin{pmatrix}\bm{x}\\\bm{y}\end{pmatrix}
+i(\bm{x}^\top\bm{u}+\bm{y}^\top\bm{v})
+\gamma\bm{v}^\top\bm{u}
\right)\!
\prod_{t=0}^T\!dx^t dy^t du^t dv^t
\right]^{\!\alpha N}\cr
&=\!\left[
\frac{1}{(2\pi)^{(T+1)}}\!\!
\int\!
\exp\!\left(
-\tfrac12(\bm{x}^\top \ \bm{y}^\top)
\Sigma\!
\begin{pmatrix}\bm{x}\\\bm{y}\end{pmatrix}
+i\bm{x}^\top(\bm{u}-\gamma S\bm{u})
+\tfrac{\gamma^2}{2}\bm{u}^\top R\bm{u}
\right)\!
\prod_{t=0}^T\!dx^t du^t
\right]^{\!\alpha N}\cr
&=\!
\left[\det Q
\det\!\left((I-\gamma S)^\top Q^{-1}(I-\gamma S)-\gamma^2 R\right)
\right]^{-\frac{\alpha N}{2}}.
\end{align}
\end{small}
where we denote 
$\bm{u}=(u^0,\ldots,u^T)^\top$, 
$\bm{v}=(v^0,\ldots,v^T)^\top$, 
$\bm{x}=(x^0,\ldots,x^T)^\top$, and 
$\bm{y}=(y^0,\ldots,y^T)^\top$.

We then substitute Eq. (\ref{eq:exp_gamma_uv}) and 
the following identities, 
\begin{align}
    1 &= N\int \delta\left (\sum_{i=1}^N
    f(a_i^t)f(a_i^s) - NQ(t,s) \right) dQ(t,s) \cr
    & =
    \frac{N}{2\pi} \int \exp\left [-
    \hat{Q}(t,s)
    \left (\sum_{i=1}^N
    f(a_i^t)f(a_i^s) - NQ(t,s) \right) 
    \right ]dQ(t,s)d\hat{Q}(t,s) 
\end{align}
and similarly for $R(t,s)$ and $S(t,s)$
for $\forall{t}, \forall{s}\in \{0,\ldots,T\}$,
into the right-hand side of Eq. (\ref{eq:mean_Z}). 
{The normalization property 
$\lim_{\tilde{\bm{\theta}} \to \bm{0}}Z(\tilde{\bm{\theta}})=1$ of Eq.~(\ref{eq:Z})
for the addition of any external inputs in Eq.~(\ref{eq:discrete_non_monotonic})
ensures $R=0$. 
} Evaluating integrals over $Q$, $R$, and $S$ using the saddle point method yields
\begin{align}
    &\hat{Q} = 0, \label{eq:hatQ} \\
    &\hat{R} = \alpha \gamma^2 (I-\gamma S)^{-1}Q
    ((I-\gamma S)^\top )^{-1}, \label{eq:hatR} \\
    &\hat{S} = \alpha \gamma \left [I + \gamma S(I-\gamma S)^{-1} \right ]. \label{eq:hatS} 
\end{align}

These results indicate that, for $N\gg 1$, Eq. (\ref{eq:mean_Z}) 
can be written as 
\begin{small}
\begin{align}
\label{eq:apdx_Z}
\mathbb{E}_{\tilde{\bm{\xi}}}[Z(\tilde{\bm{\theta}})]
\simeq&
\mathop{\rm tr}_{\bm{\sigma}^0,\tilde{\bm{a}},\tilde{\bm{h}}}
\!\left\{
P(\bm{\sigma}^0)
\exp\!\left[
-\!\sum_{t=0}^T\sum_{i=1}^N\!
ih_i^t(a_i^{t+1}-a_i^t-\gamma(-a_i^t+m^t-\alpha f(a_i^t)))
\right]
\right.\cr
&\times\!
\exp\!\left(
-\!\sum_{i=1}^N\!\sum_{t,s}\!\left[
\tfrac12\hat{R}(t,s)(ih_i^t)(ih_i^s)
+\hat{S}(t,s)f(a_i^t)(ih_i^s)
\right]\!
\right)\!
\left.\times
\exp\!\left(i \sum_{t=0}^{T+1}\bm{\theta}^t\!\cdot\!\bm{a}^t\right)
\right\}\cr
=&\,
\mathop{\rm tr}_{\bm{\sigma}^0,\tilde{\bm{a}},\tilde{\bm{h}},\tilde{\bm{\phi}}}
\!\left\{
P(\bm{\sigma}^0)
\exp\!\left[
-\!\sum_{t=0}^T\sum_{i=1}^N ih_i^t{\mathcal L}_i^t
\right]
\times\!
\exp\!\left(
i \sum_{t=0}^{T+1}
\bm{\theta}^t\!\cdot\!\bm{a}^t\right)
\right.\cr
&\left.\times
\frac{1}{(2\pi)^{N(T+1)/2}\!\det(\gamma^{-2}\hat{R})^{N/2}}
\exp\!\left(
-\tfrac{\gamma^2}{2}\!\sum_{i=1}^N\!\sum_{t,s}\!
\hat{R}^{-1}(t,s)\phi_i^t\phi_i^s
\right)
\right\}.
\end{align}
\end{small}
where 
\begin{align}
\label{eq:apdx_traj}
    {\mathcal L}_i^t = a_i^{t+1} -a_i^t 
    -\gamma\left (-a_i^t + m^t + \phi_i^t 
    -\alpha f(a_i^t) + \gamma^{-1}\sum_{s=0}^T 
    \hat{S}(t,s) f(a_i^s) \right ). 
\end{align}
Eq. (\ref{eq:apdx_Z}) implies that $S(t,s)$ can be 
assessed as
\begin{align}
\label{eq:apdx_S}
    S(t,s) &= \frac{1}{N}\sum_{i=1}^N f(a_i^t) (ih_i^s) \cr
    &\simeq\frac{1}{N}\sum_{i=1}^N \mathbb{E}_{\tilde{\bm{\phi}}}
    \left [f(a_i^t) (ih_i^s) \right ] \cr
    &= \frac{1}{N}\sum_{i=1}^N\frac{1}{\gamma} 
    \mathbb{E}_{\tilde{\bm{\phi}}}
    \left [\frac{\partial f(a_i^t)}{\partial \phi_i^s} \right ] \cr
    &= 
    \begin{cases}
    \displaystyle
    \frac{1}{N}\sum_{i=1}^N\frac{1}{\gamma} 
    \mathbb{E}_{\tilde{\bm{\phi}}}
    \left [\frac{\partial f(a_i^t)}{\partial \phi_i^s} \right ], & s < t, \cr
    0, & \mbox{otherwise}
    \end{cases}
     \cr
    &\simeq  
    \begin{cases}
    \displaystyle
    \sum_{w < t}
    \left (\frac{1}{N}\sum_{i=1}^N\frac{1}{\gamma} \left [
    f(a_i^t) \phi_i^w\right ] \right )\gamma^2 \hat{R}^{-1}(w,s),  & 
    s< t, \cr
    0, &\mbox{otherwise}, 
    \end{cases}
\end{align}
where 
we used the law of large numbers in the tranformations  
connected by $\simeq$ and a formula 
{
\begin{align}
\label{eq:Response_derivation}
    \mathbb{E}_{\tilde{\bm{\phi}}}
    \left [ \frac{\partial f(a_i^t)}{\partial \phi_i^s} \right] &= 
    \int \prod_{u=0}^{T}d\phi_i^u
    \frac{\exp\!\left(
-\tfrac{\gamma^2}{2}\!\sum_{t,s}\!
\hat{R}^{-1}(t,s)\phi_i^t\phi_i^s
\right)}{(2\pi)^{(T+1)/2}\!\det(\gamma^{-2}\hat{R})^{1/2}}
\frac{\partial f(a_i^t)}{\partial \phi_i^s} \cr
&=\int \prod_{u=0}^{T}d\phi_i^u
    \frac{\exp\!\left(
-\tfrac{\gamma^2}{2}\!\sum_{t,s}\!
\hat{R}^{-1}(t,s)\phi_i^t\phi_i^s
\right)}{(2\pi)^{(T+1)/2}\!\det(\gamma^{-2}\hat{R})^{1/2}}
f(a_i^t) \left (\sum_{w} \phi_i^w \gamma^2 \hat{R}^{-1}(w,s) \right )\cr
&=\mathbb{E}_{\tilde{\bm{\phi}}}
    \left [  f(a_i^t)\left (\sum_{w} \phi_i^w \gamma^2 \hat{R}^{-1}(w,s) \right )\right], 
\end{align}}
which is derived using integral by parts for Gaussian integral.
Between the third and fourth lines, 
we also used $\partial f(a_i^t)/\partial \phi_i^s = 0$
for $s\ge t$ due to the causality. 
Eq. (\ref{eq:apdx_S}) indicates that $S$ is a lower triangular 
matrix. As a result, $\hat{S}$ is also a lower triangular matrix, and therefore the summation with respect to $s$ 
on the right-hand side of Eq. (\ref{eq:apdx_traj}) is limited to the 
range of $0\le s \le t$. 
Combining these and setting 
\begin{align}
    G(t,s) &= \gamma S(t,s), \\
    C(t,s) & = \gamma^{-2}\hat{R}(t,s), \\
    \Lambda(t,s) & = -\alpha I + \gamma^{-1}\hat{S}(t,s) \label{eq:Gamma_final}, 
\end{align}
provide Eqs. (\ref{eq:effective_Z})--(\ref{eq:Gamma}). 
Note that Eq. (\ref{eq:Gamma_final}) indicates 
$\Lambda(t,t) = 0$. Accordingly, the summation range of $s$ in the right-hand side of Eq.~(\ref{eq:Langevin}) is written as 
$0 \le s \le t-1$. 

\section{Derivation of $\Lambda(t,s) $ at the fixed point}
\label{appendixLambda}
We assume that $C(t,s) = \sigma^2$ and $K(t,s) = K^*$ 
hold for all pairs of $t$ and $s$. 
For notational simplicity, we set $a = \sigma_\phi^{-2} K^*$. 
Under the assumption, Eq.~(\ref{eq:G}) indicates that, 
$G_t$ is given by  
\begin{align}
\label{eq:G_t}
    G_t = \left (
    \begin{array}{ccccc}
    0&0&0& \cdots & 0 \cr
    \frac{a}{2} & 0 & 0 &\cdots & 0 \cr
    \frac{a}{3} & \frac{a}{3} & 0 & \cdots& 0 \cr
    \vdots & \vdots & \vdots & \ddots & \vdots \cr
    \frac{a}{t} & \frac{a}{t} & \frac{a}{t} & \cdots & 0
    \end{array}
    \right ), 
\end{align}
and the feedback coefficient and its 
integrated form are defined as 
$\Lambda(t,s) = \alpha 
[G_t(I_t-G_t)^{-1}](t,s)$
and $\Lambda(t) = \sum_{s<t} \Lambda(t,s)$. 

To evaluate these quantities, we first define
\begin{align}
    H_t = (I_t-G_t)^{-1}, 
\end{align}
which satisfies
\begin{align}
\label{eq:H_t_eq}
    H_t = I_t + G_t H_t. 
\end{align}
Inserting Eq. (\ref{eq:G_t}) 
into (\ref{eq:H_t_eq}) offers
\begin{align}
    H_t(u,s) =
    \begin{cases}
    \displaystyle
      1   &  s=u, \\
      \frac{a}{u} \sum_{r=1}^{u-1} H_t(r,s), 
      & s < u,    \cr
      0, & s> u.
    \end{cases}
\end{align}
Since $H_t$ is a lower triangular matrix, 
this can be rewritten as
\begin{align}
    H_t(u,s) = 
    \begin{cases}
    \displaystyle
      1   &  s=u, \\
      \frac{a}{u} \sum_{r=s}^{u-1} H_t(r,s), 
      & s < u,    \cr
      0, & s> u.
    \end{cases}
\end{align}

For $s\le u$, let us 
define $F(u,s) = \sum_{r=s}^u H_t(r,s)$. 
This quantity satisfies 
\begin{align}
    F(u,s) &= F(u-1,s) + H_t(u,s) \cr
    &=C(u-1,s) + \frac{a}{u} F(u-1,s)\cr
    &= \left (1+\frac{a}{u} \right )F(u-1,s)
\end{align}
with initial condition $F(s,s)=1$. 
Hence, 
\begin{align}
    F(u-1,s) = \prod_{r=s+1}^{u-1}\left (
    1+ \frac{a}{r} \right ),  
\end{align}
and 
\begin{align}
    H_t(u,s) = \frac{a}{u}F(u-1,s) = \frac{a}{u} \prod_{r=s+1}^{u-1}\left (
    1+ \frac{a}{r} \right ),  
\end{align}
where we defined the empty product $\displaystyle \prod_{r=s+1}^{u-1}\left (
1+ \frac{a}{r} \right )=1$ for $s=u-1$. 

Since 
\begin{align}
    \Lambda(t,s) = \alpha [G_tH_t](t,s)
=\alpha \left ( H_t(t,s) -\delta_{t,s} \right ), 
\end{align} 
we obtain
\begin{align}
    \Lambda(u,s) &=
    \begin{cases}
    \displaystyle
      \frac{a}{t} \prod_{u=s+1}^{t-1}
      \left (1+\frac{a}{u} \right ),&  s<t, \\
      0, & s \ge t,    
    \end{cases}
     \cr
    &=
    \begin{cases}
    \displaystyle
      \frac{a}{t} \frac{\Gamma(t+a)}{\Gamma(t)}
      \frac{\Gamma(s+1)}{\Gamma(s+1+a)}, &  s<t, \\
      0, & s \ge t.    
    \end{cases}
\end{align}
Finally, using the asymptotic form of the Gamma function,
$\Gamma(t+a)/\Gamma(t) \sim t^a$ for $t \gg a$, 
we recover Eq. (\ref{eq:Lambda_ts_asympt}).

\section{Derivation of the fixed point conditions}
\label{appendix2}
By dropping the time dependence of all relevant quantities, 
the stationary (convergent) solution of Eq.~(\ref{eq:Langevin}) is characterized by
\begin{align}
\label{eq:cavity_zeroth}
    a_i &= \sum_{j\ne i} J_{ij}f(a_i) \cr
    &= \frac{1}{N} \sum_{j=1}^N f(a_i) 
    + \sum_{\mu=2}^{\alpha N} \frac{1}{N}\xi_i^\mu 
    \sum_{j=1}^N \xi_j^\mu f(a_i) -\alpha f(a_i) \cr
    & = m + \phi_i +\Lambda f(a_i). 
\end{align}
Operating $f(\cdot)$ to this yields another expression 
\begin{align}
\label{eq:cavity_first}
    s_i = f(m + \phi_i +\Lambda s_i),  
\end{align}
where we set $s_i = f(a_i)$ for the convergent solution. 

Let the solution of Eq.~(\ref{eq:cavity_first}) be expressed as
\begin{align}
\label{eq:cavity_second}
s_i = g(m + \phi_i),
\end{align}
where $g(\cdot)$ represents the effective transfer function defined implicitly by Eq.~(\ref{eq:cavity_first}).
Using this relation, the condition for the output overlap becomes
\begin{align}
\label{eq:fixed_m}
m = \frac{1}{N}\sum_{i=1}^{N}s_i
= \frac{1}{N}\sum_{i=1}^{N} g(m+\phi_i).
\end{align}

To evaluate the variance $\sigma_\phi^2$ of $\phi_i$ and the convergent value $\Lambda$ of the integrated feedback coefficient $\Lambda(t) = \sum_{s<t}\Lambda(t,s)$,
we employ the cavity method~\cite{Mezard1987}.
The random variable $\phi_i$ represents the random component of the cross-talk noise
$\sum_{\mu=2}^{\alpha N} N^{-1}\xi_i^\mu \sum_{j=1}^N \xi_j^\mu f(a_j)$
in the second line of Eq.~(\ref{eq:cavity_zeroth}).
To extract this component, we decompose the cross-talk noise as
\begin{align}
\label{eq:cross_talk_decomposition}
    \frac{1}{N} \sum_{\mu=2}^{\alpha N}
    \xi_i^\mu 
    \sum_{j=1}^N \xi_j^\mu s_j
    &= \psi_i^{\backslash \mu}  + \frac{\xi_i^\mu}{\sqrt{N}}\Delta^\mu
\end{align}    
focusing on a specific pattern $\bm{\xi}^\mu$ ($\mu \in \{2,\ldots,\alpha N\}$),
where 
\begin{align}
\label{eq:psi_mu}
    \Delta^\mu = \frac{1}{\sqrt{N}}\sum_{i=1}^N \xi_i^\mu s_i
\end{align}
and
\begin{align}
\label{eq:phi_i_mu}
    \psi_i^{\backslash \mu} &= \frac{1}{N} \sum_{\nu \ne 1,\mu}^{\alpha N}
    \xi_i^\nu 
    \sum_{j=1}^N \xi_j^\nu s_j = \sum_{\nu \ne 1, \mu}
    \frac{\xi_i^\nu}{\sqrt{N}}\Delta ^\nu. 
\end{align}

Next, we express $s_i$ in terms of the solution $s_{i\to \mu}$ of the $\mu$-cavity system—defined by removing $\bm{\xi}^\mu$ from the original system—as
\begin{align}
    s_i  &= s_{i\to \mu} + \frac{\partial s_{i}}{\partial \phi_i}
    \frac{\xi_i^\mu }{\sqrt{N}}\Delta^\mu  + O(N^{-1}) \cr
    & = s_{i\to \mu} + g^\prime(m+\phi_i) 
    \frac{\xi_i^\mu }{\sqrt{N}}\Delta^\mu  + O(N^{-1}).  
\end{align}
Substituting this into Eq. (\ref{eq:psi_mu}) offers
\begin{align}
\label{eq:psi_mu2}
    \Delta^\mu = \frac{1}{\sqrt{N}}\sum_{i=1}^N \xi_i^\mu s_{i\to \mu} + U
    \Delta^\mu + O(N^{-1/2}), 
\end{align}
and therefore we can set 
\begin{align}
\label{eq:psi_mu3}
    \Delta^\mu = \frac{1}{1-U}\times \frac{1}{\sqrt{N}}\sum_{i=1}^N \xi_i^\mu s_{i\to \mu}, 
\end{align}
where we defined 
\begin{align}
\label{eq:fixed_U}
    U = \frac{1}{N} \sum_{i=1}^N g^\prime(m+\phi_i).  
\end{align}

Since $\bm{\xi}^\mu$ and $\bm{s}^{\backslash\mu}=(s_{i\to \mu})$ are statistically independent,
the central limit theorem ensures that
$N^{-1/2}\sum_{i=1}^N\xi_i^\mu s_{i\to \mu}$ follows a zero-mean Gaussian distribution
with variance $N^{-1}\sum_{i=1}^N s_{i\to \mu}^2 \simeq 
N^{-1}\sum_{i=1}^N s_{i}^2
=
N^{-1}\sum_{i=1}^N[g(m+\phi_i)]^2$.
Combining this with Eq.~(\ref{eq:psi_mu3}) gives
\begin{align}
{\rm var}[\Delta^\mu] = \frac{1}{N} \sum_{i=1}^N \frac{[g(m+\phi_i)]^2}{(1-U)^2}.     
\end{align}

Substituting Eq.~(\ref{eq:psi_mu3}) into
Eqs.~(\ref{eq:cross_talk_decomposition}) and (\ref{eq:phi_i_mu}),
and noting that
\begin{align}
    \frac{\xi_i^\mu}{\sqrt{N}} \Delta^\mu
    &=\frac{1}{1-U} \times \frac{\xi_i^\mu }{\sqrt{N}}\sum_{j = 1}^N \xi_j^\mu 
    s_{j\to \mu} \cr
    & = \frac{1}{1-U}\left (\frac{\xi_i^\mu}{\sqrt{N}}
    \sum_{j\ne i} \xi_j^\mu s_{j\to \mu} \right ) + \frac{s_{i\to \mu}}{N(1-U)} \cr
    & = \frac{1}{1-U}\left (\frac{\xi_i^\mu}{\sqrt{N}}
    \sum_{j\ne i} \xi_j^\mu s_{j\to \mu} \right ) +
    \frac{s_i}{N(1-U)}
    + O(N^{-3/2})
\end{align}
holds for $\forall \mu \in \{2,\ldots,\alpha N\}$. 
We find that the second term in the last line, though negligible for a single $\mu$,
contributes collectively after summing over $\mu\in \{2,\ldots,\alpha N\}$.
This yields 
\begin{align}
\label{eq:cross_talk_decomposition2}
    \frac{1}{N} \sum_{\mu=2}^{\alpha N}
    \xi_i^\mu 
    \sum_{j\ne i} \xi_j^\mu s_i
    &= \frac{1}{1-U} 
    \sum_{\mu =2}^{\alpha N} 
    \frac{\xi_i^\mu}{\sqrt{N}}
    \left (
    \sum_{j \ne i}
    \frac{\xi_j^\mu s_{j\to \mu}}{\sqrt{N}}
    \right )
    + \frac{\alpha}{1-U}s_i -\alpha s_i  + O(N^{-1/2})\cr
    & = \frac{1}{1-U} 
    \sum_{\mu =2}^{\alpha N} 
    \frac{\xi_i^\mu}{\sqrt{N}}
    \left (
    \sum_{j =1}^N 
    \frac{\xi_j^\mu s_{j\to \mu}}{\sqrt{N}}
    \right )
    + \frac{\alpha U}{1-U}s_i + O(N^{-1/2}) \cr
    & = \sum_{\mu=2}^{\alpha N}\frac{\xi_i^\mu}{\sqrt{N}}
    \Delta^\mu + \frac{\alpha U}{1-U} g(m+\phi_i) + O(N^{-1/2}). 
\end{align}  
The first term corresponds to $\phi_i$, while the coefficient of the second term defines $\Lambda$.
Hence,
\begin{align}
\label{eq:fixed_sigma2}
    \sigma_\phi^2 = \sum_{\mu=2}^{\alpha N} \frac{1}{N} {\rm var}[\Delta^\mu] =\frac{\alpha}{N} \sum_{i=1}^N \frac{[g(m+\phi_i)]^2}{(1-U)^2}
\end{align}
and 
\begin{align}
\label{eq:fixed_Gamma}
    \Lambda= \frac{\alpha U}{1-U}. 
\end{align}
Replacing $N$ with $N_{\rm DMFT}$ in
Eqs.~(\ref{eq:fixed_m}), (\ref{eq:fixed_U}), (\ref{eq:fixed_sigma2}), and (\ref{eq:fixed_Gamma})
yields the fixed-point conditions
Eqs.~(\ref{eq:m_macro})--(\ref{eq:Gamma_macro}).


\providecommand{\noopsort}[1]{}\providecommand{\singleletter}[1]{#1}%

\end{document}